\documentclass[twocolumn]{aastex631}

\usepackage{url}
\usepackage[caption=false]{subfig}
\usepackage{mathtools}
\let\tablenum\relax
\usepackage{siunitx}
\usepackage{xspace}
\usepackage{threeparttable} 
\begin{document}

\title{Long-term monitoring of repeating FRB 20220912A with the uGMRT at low radio frequencies}

\author[0009-0002-0330-9188]{Ajay Kumar}
\email{akumar@ncra.tifr.res.in}
\affiliation{National Centre for Radio Astrophysics (NCRA - TIFR), Pune - 411007, India}

\author[0000-0002-0862-6062]{Yogesh Maan}
\email{ymaan@ncra.tifr.res.in}
\affiliation{National Centre for Radio Astrophysics (NCRA - TIFR), Pune - 411007, India}

\author[0000-0002-0862-6062]{Banshi Lal}
\affiliation{National Centre for Radio Astrophysics (NCRA - TIFR), Pune - 411007, India}

\author[0000-0002-5342-163X]{Yash Bhusare}
\affiliation{National Centre for Radio Astrophysics (NCRA - TIFR), Pune - 411007, India}

\author[0000-0002-5342-163X]{Shriharsh P. Tendulkar}
\affiliation{National Centre for Radio Astrophysics (NCRA - TIFR), Pune - 411007, India}
\affiliation{Tata Institute for Fundamental Research, Mumbai, India}

\author[0000-0000-0000-0000]{Visweshwar Ram Marthi}
\affiliation{National Centre for Radio Astrophysics (NCRA - TIFR), Pune - 411007, India}

\author[0000-0002-2711-2366]{Puja Majee}
\affiliation{National Centre for Radio Astrophysics (NCRA - TIFR), Pune - 411007, India}

\begin{abstract}
Some repeating FRBs exhibit occasional extreme repetition rates, but very few show a sustained high activity level. One such hyperactive repeater is FRB\,20220912A, which was discovered by CHIME/FRB Collaboration on 2022 September 12. Here, we present results from a long-term monitoring campaign of FRB\,20220912A using the upgraded Giant Metrewave Radio Telescope (uGMRT) in the frequency range from 300 to 750 MHz. Over the course of nearly two years, we detected a total of 643 bursts in this frequency range. The source exhibited extreme activity for a few months after its discovery and sustained its active phase for more than 1.5\,years, with unsystematic modulations in the activity during this phase. The cumulative energy distributions in both bands show a break, consistent with other active repeaters like FRB\,20121102A, FRB\,202011124A, etc., suggesting common underlying emission mechanisms. Moreover, we show that the energy distribution shape for FRB\,20220912A remains broadly same across a large range of frequencies and over time. Overall, the extended high activity, estimated total energy output, persistent power-law tails in the energy distributions, and the lack of detectable short timescale periodicity favor progenitor models invoking young dynamic magnetars, potentially emitting pulses across large rotation phase ranges. 
\end{abstract}

\section{Introduction} \label{sec:intro}
FRBs are micro- to millisecond-duration coherent radio bursts in the sky that are bright ($\sim$Jy) events of cosmological origin. Currently, about 1000 FRBs have been published to date, which is expected to grow to several thousands with upcoming and ongoing dedicated and commensal FRB surveys. A small fraction of the total FRBs are known to repeat. However, their sources of origins still remain unclear. Their extragalactic nature makes them excellent probes for studying a range of astrophysical phenomena, e.g., the missing baryon problem, testing Lorentz invariance, etc \citep{Macquart2020,Vardanyan_2023}.
\par
There is still a debate about whether all FRBs repeat, since we may not have detected repeat faint bursts from the apparently non-repeating FRBs due to repeat bursts being too faint or the intrinsic rate being very low \citep{Lin_2024}. Many repeating FRBs are well localized and have identified hosts \citep{Chatterjee2017,2019Sci...365..565B,2020Natur.577..190M,2020ApJ...895L..37B,2020ApJ...903..152H}. Repeating FRBs have been localised to different types of host environments from star-forming regions to globular clusters \citep{bhardwaj2021_m81,gordon2023_host_galaxy}, which suggests different formation channels for FRBs. Persistent Radio Sources (PRS) have also been associated with FRB\,20121102A, FRB\,20190520B, FRB 20201124A, FRB\,20240114A, and FRB\,20190417A which has raised the question if there could be separate classes of FRBs \citep{2022casey}. Periodicity of 16.35 days has been seen in repeating FRB\,20180916B \citep{2020Natur.582..351C} and putative period of 157 days in FRB\,20121102A \citep{Rajwade_2020}, however, other extreme repeaters like FRB\,20190520B and FRB\,20201124A do not show periodicity in their activity despite extensive monitoring \citep{Lanman_2022,Niu_2022}. Chromaticity was seen in the activity of the FRB\,20180916B \citep{pluenis2021lofar,ines2021}. A handful of repeating FRBs show extremely high activity, i.e., burst rates $\geq$ 100 hr$^{-1}$. Extreme activities sustained over several days have been shown by only a few repeating FRBs, contributing significantly to the all-sky rate of FRBs \citep{kirsten2023connecting,omar2024_r117}. This subset also provides large samples of bursts to do robust statistical analysis of burst properties and understand the underlying emission mechanism for active repeating FRBs \citep{totani2023_earthquake_resemblance}.  
\par
Brightest bursts observed from SGR 1935+2154 are still a couple of orders of magnitude short of the typical luminosity of the extragalactic FRB population. However, recent observations of intense radio bursts from SGR 1935+2154 have shown that magnetars are capable of producing radio emission across a broad range of luminosities \citep{Kirsten_2020}. Recently, S-shaped polarisation position angle (PA) swing has been seen for a non-repeating FRB similar to canonical radio pulsars, providing further evidence towards a neutron star origin for FRBs \citep{mckinven2024pulsarlikeswingpolarisationposition}.    

On 2022 September 12, CHIME discovered FRB\,20220912A, detecting nine bursts from it in three days at a nominal DM of $219.46$ pc cm$^{-3}$ \citep{2022ATel15679....1M}. Shortly after, it was localised with arcsec precision to RA=23h09m04.9s Dec=48d42'25.4" with the DSA-110 telescope, and the identified host galaxy has a redshift of $0.771$ \citep{Ravi_2023}. Prompt follow-ups of FRB\,20220912A detected extremely high activity for the source, with detections up to $2.3$ GHz \citep{2022ATel15723....1F,2022ATel15727....1K,2022ATel15733....1Z,2022ATel15806....1B,2022ATel15734....1P}. A majority of the observed bursts show narrowband spectra \citep{Zhang2023r117fast} typical of repeating FRBs, and some bursts are very narrow in time and clustered in dense burst forests \citep{Hewitt_2023}. Physical origins of characteristic narrowband spectra for repeating FRBs have also been explored and related to magnetar models \citep{2023ApJ...956...67Y}. \citet{Feng2023_r117gbt} concluded that the source resides in a cleaner environment in contrast to some of the active repeaters \citep{Mckinven_2023,Xu_comlexenvr67_2022} which is supported by the fact that the DM contribution by the host is low ($\leq53$ pc cm$^{-3}$) and RM measurement of $\sim$\,0\,rad m$^{-2}$ \citep{Ravi_2023,Feng2023_r117gbt, Zhang2023r117fast, Hewitt_2023}. A trend or evolution is also seen in the central frequency of band-limited bursts from FRB\,20220912A over time \citep{Sheik2024r117Allen} and a secular increase in DM is also reported over time \citep{abbott2026radiomonitoringcampaignactive}. \citet{pelliciari2024_northencross} reported non-detection at X-ray and $\gamma$-ray wavelengths with simultaneous observations with radio using Swift and AGILE, respectively. 
\par
Long-term monitoring of an active repeating FRB can help us in studying the temporal variation of burst rate, spectro-temporal, and polarization properties. Low-frequency radio studies can reveal certain properties of FRBs like scattering, scintillation, chromaticity in the activity \citep{ines2021,pluenis2021lofar}, spectral index, etc. Several repeaters have shown evolution in their burst properties and activity levels with time. Upgraded GMRT (uGMRT; \citep{gupta_2017}) is an excellent observational facility to monitor highly active sources owing to its higher sensitivity ($\sim$ 10 times more sensitive than CHIME (400-800MHz) and wider low-frequency coverage with dual array configuration in band-3 (300-500 MHz) and band-4 (550-750 MHz). Following the CHIME discovery of FRB\,20220912A, we embarked upon a long-term, multi-frequency monitoring of this source. In this paper, we report the details of our search method, completeness analysis, first-ever detections of bursts from FRB\,20220912A and their energy distributions at two different frequencies obtained from long-term monitoring GMRT campaign. Other properties such as waiting time distribution, structure-optimised DM, emission bandwidth, scatter-broadening and scintillation will be reported separately in Bhusare et al. (in prep) and Majee et al. (in prep). We describe the observation setups and search methods in Section~\ref{sec:observations} and present results from our analysis in Section~\ref{sec:results}. We discuss the implications of the results in the context of repeating FRBs in section~\ref{sec:discussion_r117} and summarize in Section~\ref{sec:conclusion}.

\section{Observations and search methods}    \label{sec:observations}
GMRT is a Y-shaped radio interferometer with 30 dishes. All our observations are conducted in phased array mode, where we use 22 dishes, leaving out the outer antennas of the three arms to keep the array phased for a longer duration. Within this phased array framework, we employ two different array configurations in different observations. In the single-array (SA) setup, all 22 antennas are used together at a single band, i.e., either at band-3 (300–500 MHz) or band-4 (550–750 MHz), with the observations at the two bands performed sequentially, one after the other. In the dual-array (DA) setup, the 22 antennas are divided into two sub-arrays, where half of the antennas are used at band-3 and the other half at band-4, allowing simultaneous dual-band coverage from 300 to 750\,MHz\footnote{Note: There is a gap of 50 MHz from 500 MHz to 550 MHz.}.
Our observing campaign of FRB 20220912A spanned from 2022 November 22 to 2024 August 24, with a non-uniform cadence due to scheduling constraints and varying time allocations. The initial observations on 2022 November 22 were conducted in band-3 and band-4 in the single-array setup. From 2022 November 24, we transitioned to the dual-array setup to achieve simultaneous wide-frequency coverage from 300 to 750\,MHz, continuing until 2023 September 24. These observations confirmed the high burst activity of FRB 20220912A at low radio frequencies \citep{2022ATel15806....1B}. Thereafter, we reverted to the single-array setup, observing at band-3 and band-4 sequentially. Additionally, in three observing sessions between MJD 60377 and 60401, we extended our frequency coverage to band-5 (1060–1460 MHz) in the single-array setup. Note that at some of the epochs, the number of antennas used in each band varied due to the unavailability of some antennas for technical or maintenance reasons. The details of all observations are provided in Table~\ref{tab:my-table}.

\par
On 2022 November 22, we recorded phased array (PA) beam data with 8192 frequency channels and 655.36 $\mu$s sampling time in band-3, and 4096 frequency channels and 327.68 $\mu$s sampling time in band-4. Later, in each observation, we recorded PA beam data with a sampling time of 327.68 $\mu$s and frequency resolution of 48.82 kHz in both band-3 and band-4. For band-5, we recorded with 2048 frequency channels and 163.94 $\mu$s sampling time. For some later observations in the campaign, we recorded the post-correlation \citep[PC;][]{Roy_2018} beam data instead of the PA beam, which typically has fewer RFI events and lesser baseline fluctuations compared to the PA beam data. We also recorded interferometric visibilities with a 10.7 sec integration time for each session in both bands, and results from those have been reported in \citet{bhusare2024lowfrequencyprobespersistentradio}. In each observation, we utilise the simultaneous recording of Phased Array Spectral Voltage \citep[PASV;][]{Marthi_2024} data in both bands, which allows offline processing to obtain coherently dedispersed high-time and frequency resolution phased-array data.
\par
Initially, we applied RFIClean\footnote{https://github.com/ymaan4/RFIClean} \citep{Maan_2021} to mitigate periodic, broadband, and spiky radio frequency interference (RFI) and to convert the data into SIGPROC filterbank format. Following this, the \texttt{dedisperse} utility from \texttt{SIGPROC} was used to divide the data into 1024 frequency sub-bands, after dedispersing the channels within each sub-band. A second round of RFI mitigation was then performed using RFIClean. Finally, the \texttt{rfifind} tool from \texttt{PRESTO} \citep{2011ascl.soft07017R} was employed to identify additional RFI-contaminated segments and generate an RFI mask. After applying this mask to the filterbank file, de-dispersion is performed on this file. 
\par
For dedispersion, we utilised \texttt{prepdata} from \texttt{PRESTO} to search over a dispersion measure (DM) range of 215 to 230 pc cm$^{-3}$ with steps of $0.1$ pc cm$^{-3}$ for band-3 and 0.2 pc cm$^{-3}$ for band-4. The resulting de-dispersed time series were used to search for single pulses using \texttt{single\_pulse\_search.py}, applying a signal-to-noise threshold of 7 and a maximum boxcar width of 0.5\,sec. Detected candidates were clustered based on their arrival times and DM values, with the highest signal-to-noise ratio candidate from each cluster was visually inspected using waterfall plots and DM-time plots generated by \texttt{your} package \citep{Aggarwal2020}. In total, we detect 643 bursts across all epochs in the observing campaign.
\par
For observations done on 2022 November 24 and 22, we also perform a sub-band search to detect bursts exhibiting narrowband emission. To perform the sub-band search, we divide the 200 MHz bandwidth in both band-4 and band-3 into four sub-bands. Each sub-band covers a 50 MHz bandwidth, with central frequencies ranging from 325 to 475 MHz in band-3 and from 575 to 725 MHz in band-4, incremented by 50 MHz. For de-dispersion, we exclude those frequency channels that fall outside the desired sub-band. We use a DM step of 1 pc cm$^{-3}$ for both bands. The rest of the search process follows the same method as previously described. However, we did not find any additional bursts with the sub-band search. 


\begin{table}[]
\caption{Details of the observations}
\centering
\begin{tabular}{l|l|l|l|l|l} \hline
Start UTC              & MJD                & $\nu_{obs}$ & T$_{obs}^{*}$ & N$^{\#}$ & Setup \\ 
Date and Time$^{\dag}$ &                    & MHz       &  min             &              &       \\ \hline
22/11/2022 16:58       & 59905.71           & 300-500     & 28.6          & 57           & SA    \\
22/11/2022 17:56       & 59905.75           & 550-750     & 35.0          & 57           & SA    \\
24/11/2022 17:36       & 59907.73           & 300-750     & 78.8          & 164          & DA    \\
30/01/2023 11:13       & 59974.47           & 300-750     & 60.2          & 4            & DA    \\
25/02/2023 10:37       & 60000.44           & 300-750     & 60.4          & 1            & DA    \\
20/03/2023 06:03       & 60023.25           & 300-750     & 74.0          & 29           & DA    \\
01/05/2023 06:28       & 60065.27           & 300-750     & 48.3          & 57           & DA    \\
02/06/2023 01:18       & 60097.05           & 300-750     & 74.0          & 37           & DA    \\
02/07/2023 21:47       & 60126.91           & 300-750     & 74.1          & 52           & DA    \\
14/07/2023 22:32       & 60139.94           & 300-750     & 74.0          & 31           & DA    \\
08/08/2023 21:32       & 60164.90           & 300-750     & 68.6          & 105          & DA    \\
10/08/2023 19:21       & 60166.81           & 300-750     & 55.1          & 42           & DA    \\
01/09/2023 00:27       & 60188.02           & 300-750     & 36.9          & 1            & DA    \\
12/09/2023 16:27       & 60199.68           & 300-750     & 43.9          & 2            & DA    \\
24/12/2023 13:22       & 60211.56           & 300-750     & 48.6          & 4            & DA    \\
08/03/2024 06:58       & 60377.33           & 300-1460    & 99.0          & 0            & SA    \\
18/03/2024 08:02       & 60387.37           & 300-1460    & 95.0          & 0            & SA    \\
31/03/2024 06:46       & 60400.33           & 300-1460    & 103.0         & 0            & SA    \\
29/04/2024 22:47       & 60429.98           & 300-750     & 71.0          & 0            & SA    \\
27/05/2024 22:25       & 60457.93           & 300-750     & 51.0          & 0            & SA    \\
29/06/2024 02:00       & 60490.06           & 300-750     & 166.0         & 0            & SA    \\
28/07/2024 02:10       & 60519.06           & 300-750     & 66.0          & 0            & SA    \\
27/08/2024 19:58       & 60549.81           & 300-750     & 64.0          & 0            & SA    \\
27/09/2024 19:08       & 60580.81           & 300-750     & 22.0          & 0            & SA    \\ \hline
\end{tabular}
\begin{tablenotes}
\item $^{\dag}$ Start time of observations.
\item $^{*}$ Total \emph{on-source} integration time. This time does not include any gaps between scans or the time spent on phasing of the array and overheads.
\item $^{\#}$ Number of bursts detected.
\item SA: Single Array, DA: Dual Array.
\end{tablenotes}
\label{tab:my-table}
\end{table}
%

\section{Analysis and Results} \label{sec:results} 
A total of 57 bursts were detected separately in each of the band-3 and band-4 observations on 2022 November 22, i.e., a total of 114 bursts. On 2022 November 24, we carried out a simultaneous observation in both band-3 and band-4 lasting 78.8 minutes and detected 164 bursts in total, 74 in band-3 and 90 in band-4. Such high burst rates for a repeating FRB were recorded with the GMRT for the first time after FRB\,20201124A \citep{marthi2022}. Observations at L-band with FAST and GBT showed similar high burst rates \citep{2022ATel15723....1F,2022ATel15733....1Z}.
\par
\citet{Zhang2023r117fast} show that the burst rate at L-band fell sharply in 2022 December after being highly active for more than two months after its discovery on 2022 September 12 (see bottom panel of their Figure 1). Our next GMRT observations could only be scheduled on 2023 January 30 so we cannot determine whether the source became highly active again in January. However, on 2023 January 30, we measured that the burst rate had already declined significantly compared to 2022 November 24. The variation of the burst rate over the whole observing campaign till 2024 September 28 is described in section~\ref{sec:activity_variation}.
\par
\begin{figure*}
    \centering
    \includegraphics[ width=0.85\textwidth, trim=0.2cm 0.2cm 0.2cm 0.2cm, clip]{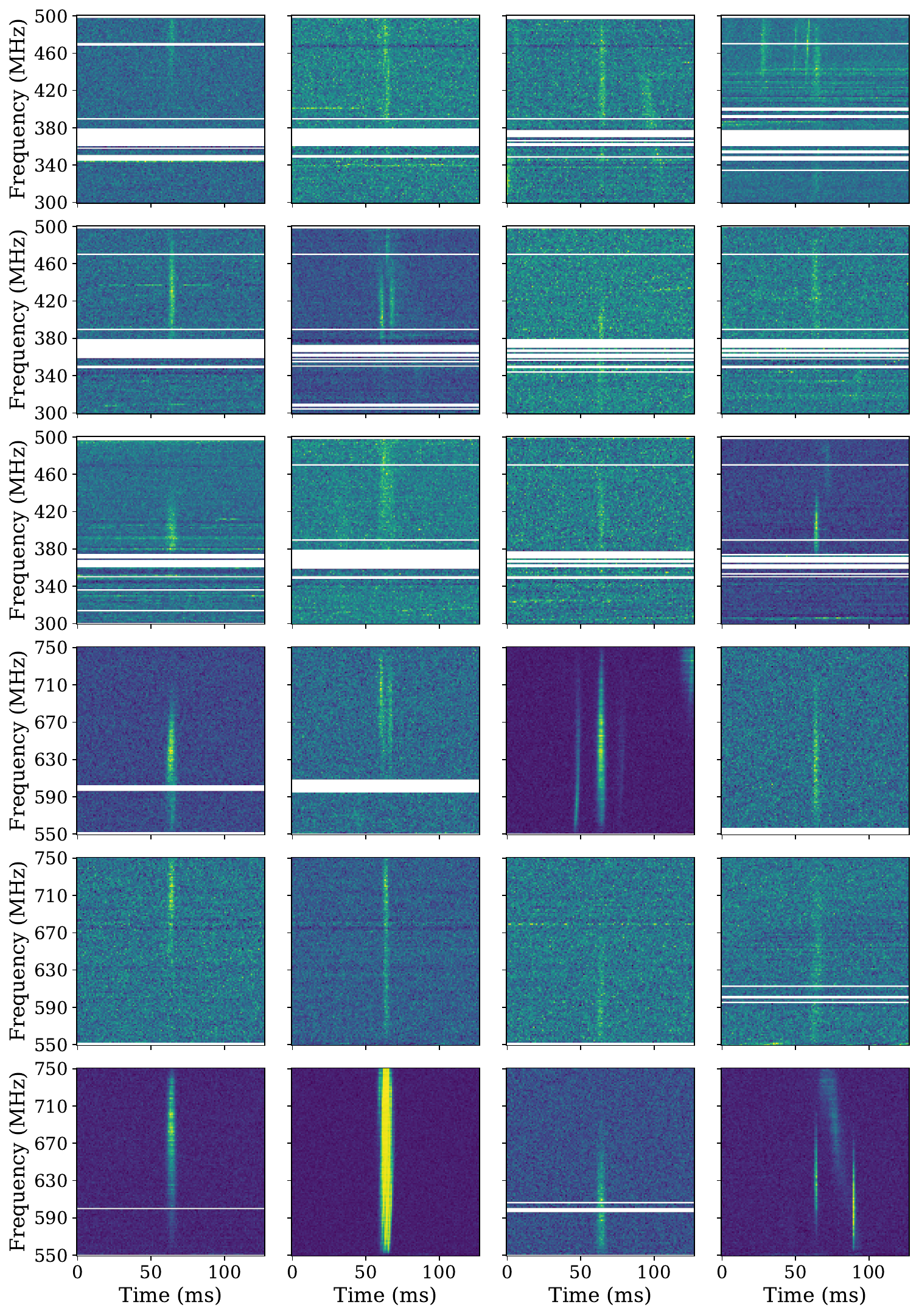}
    \caption{ Dynamic spectra of some of the bursts detected during the observing campaign in Band-3 (300–500 MHz) and Band-4 (550–750 MHz). Each dynamic spectrum comprises 128 frequency sub-bands, each 6.25\,MHz wide, and a time resolution of 2.621 ms. While sub-banding, a sub-band is fully masked if more than 50\% of the channels in it are found to be RFI contaminated.}
    \label{fig:dynamicspectra_examplebursts}
\end{figure*}

We also detected a few bursts simultaneously in band-3 and band-4. At our highest time resolution, some bursts reveal complex internal structure. Some bursts observed during the observing campaign in both band-3 and band-4 are shown in Figure~\ref{fig:dynamicspectra_examplebursts}, which show narrow-band emission and downward drifting pattern, typical of repeating FRBs.
 
\subsection{Fluence measurements}  \label{sec:fluence_measure}
We determine the fluence of each burst by utilising the calibrated band-averaged time profile. First, the time samples surrounding the burst arrival time are extracted and normalised by the RMS value measured from the off-pulse region. The time series is calibrated using the system-equivalent flux density (SEFD) for each epoch. SEFD as a function of frequency was obtained using the polynomial fit provided in the GMRT Exposure Time Calculator\footnote{\url{http://www.ncra.tifr.res.in/~secr-ops/etc/etc.html}}. A SEFD for the whole band was then computed by averaging the frequency-dependent SEFD across the band. SEFD varies for each epoch due to the different number of antennas used. Finally, we calculate the fluence by summing the area under the calibrated burst profile.

\subsection{Completeness Analysis}   \label{sec:completeness} 
To draw meaningful inferences from the burst rates and the energy distribution of the detected bursts, it is essential to first determine the completeness of the sample. To accurately estimate the completeness thresholds at both band-3 and band-4, we inject simulated dispersed pulses into the native beamformed data recorded at the observatory. For this purpose, we use the data recorded on 2022 November 24 to inject simulated bursts at a DM of 235\,pc\,cm$^{-3}$. We chose this DM value since it is closer to the FRB’s true DM but far enough that our search for simulated bursts is not affected by the real bursts. 

Our injection code uses \texttt{simpulse}\,\footnote{\url{https://github.com/kmsmith137/simpulse} } to simulate dispersed pulses in both frequency and time. Each injection is assigned a DM, an intrinsic width, a fluence (in arbitrary units), a scattering timescale, and a spectral index, as well as an arrival time. The difference in arrival time of two successive bursts is chosen such that there is no overlap between them. For each injection run, we keep the intrinsic width the same. For simplicity, we set the spectral index to zero, and we assume negligible scatter-broadening (scattering timescale is set to zero).
\par
\begin{figure*}
    \centering
    \includegraphics[width=0.45\textwidth]{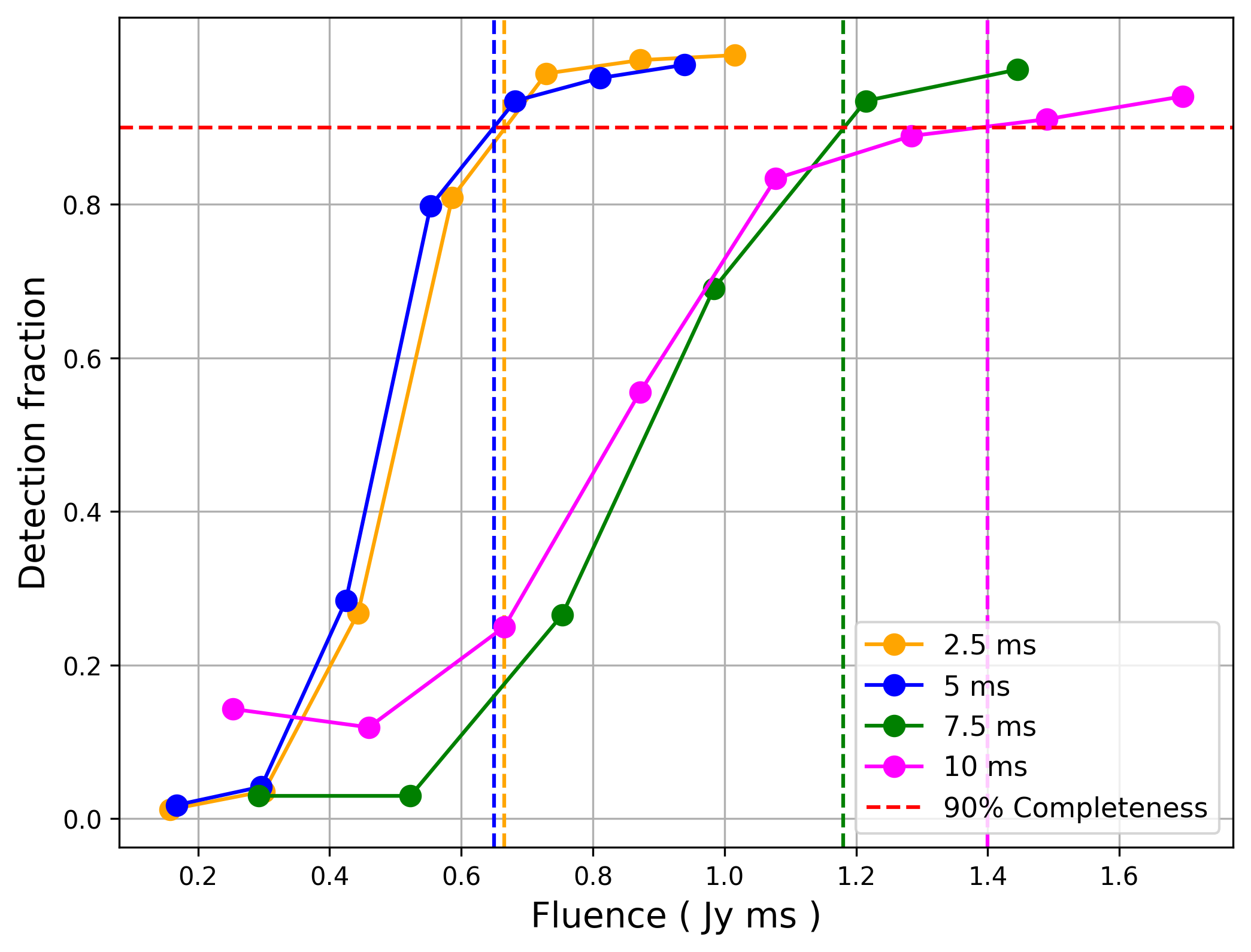}
    \hfill
    \includegraphics[width=0.45\textwidth]{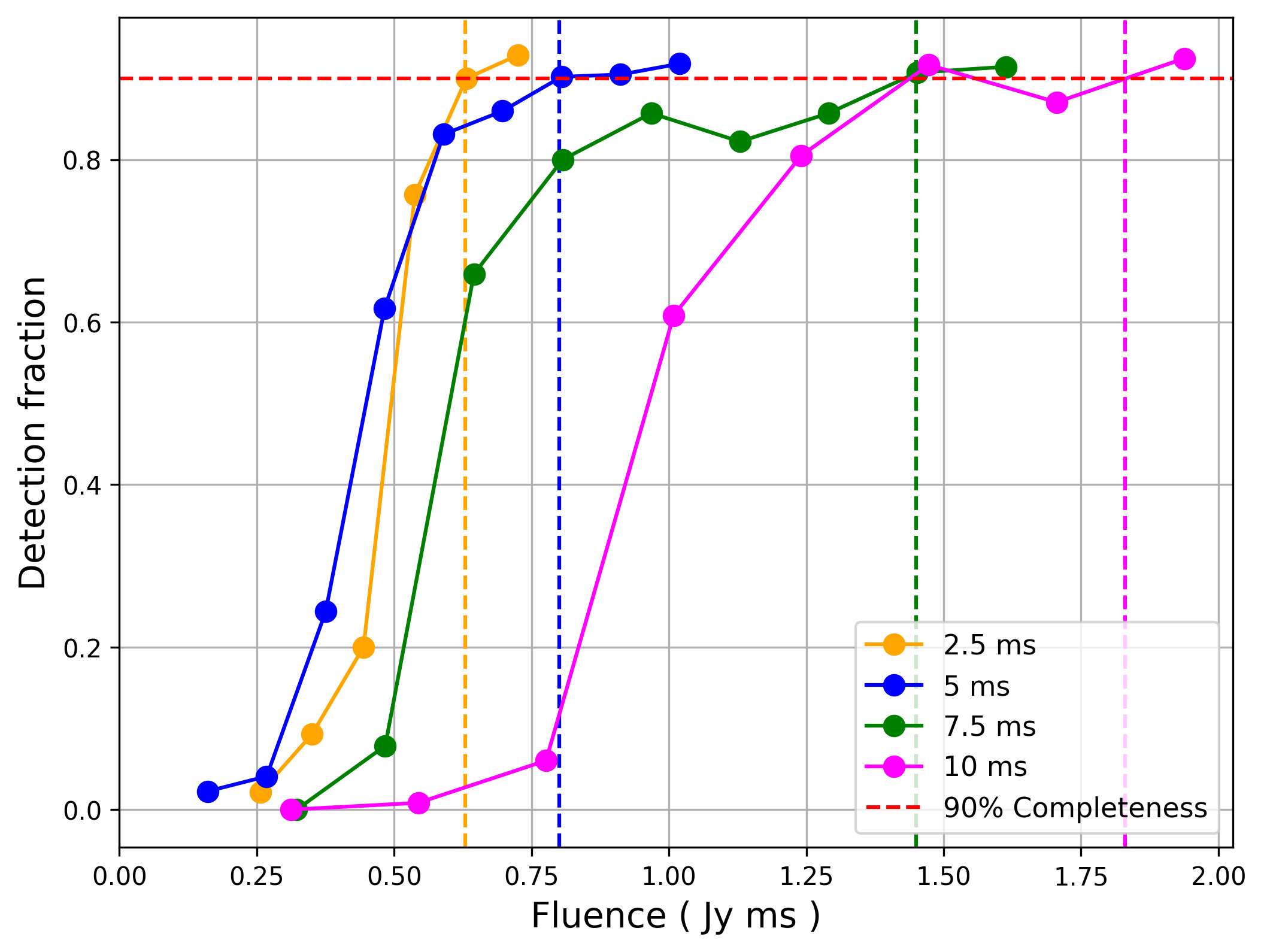}
    \caption{Completeness function for band-3 (left panel) and band-4 (right panel) at different widths, determined by injections of bursts in raw unprocessed data taken on 2022 November 24.}
    \label{fig:completeness_function}  
\end{figure*}

Because the detection significance of a pulse depends on its width, we determine completeness thresholds for widths of 2.5\,ms, 5\,ms, 7.5\,ms, and 10\,ms (this range cover most of the detected bursts). The fluence value for the pulse generated from \texttt{simpulse} is in arbitrary units. To calibrate, we first determine the median signal-to-noise ratio (S/N; from PRESTO search) of 100 bursts injected at a particular fluence value. Comparing this median S/N to the known S/N from \texttt{simpulse} yields a scaling factor.  We determine this factor for each particular width, separately for both band-3 and band-4. Using this scaling factor and the system equivalent flux density (SEFD) of the telescope, we convert \texttt{simpulse}’s arbitrary fluence units into physical units (Jy\,ms). We inject bursts uniformly distributed over the range of fluence values shown in Figure~\ref{fig:completeness_function} for both band-3 and band-4 at each particular width.

\par
We inject 1250 bursts for each width to compute robust completeness functions at different widths, for both band-3 and band-4. The left panel in Figure~\ref{fig:completeness_function} shows the completeness functions for band-3 at different widths. The 90\% completeness thresholds for 2.5, 5, 7.5 and 10\,ms widths are 0.67, 0.65, 1.18 and 1.4\,Jy ms, and the corresponding flux densities would imply theoretical detection significances of about 9$\sigma$, 7$\sigma$, 10$\sigma$, and 10$\sigma$, respectively. The right panel in Figure~\ref{fig:completeness_function} shows the completeness function for band-4. The completeness threshold for 2.5, 5, 7.5 and 10\,ms widths are 0.63, 0.8, 1.45 and 1.83\,Jy\,ms, and the corresponding flux densities would imply theoretical detection significances of about 8$\sigma$, 8$\sigma$, 12$\sigma$, and 13$\sigma$, respectively. 
\par
We note that our completeness analysis utilizes only the frequency-averaged bursts to estimate the detection significances, and is agnostic to the narrow-band or wide-band nature of the bursts. However, our approach is consistent with the subsequent energetics analysis (see Section~\ref{sec:energetics}) as all the fluence estimates are derived from the frequency-averaged burst profiles.


\subsection{Temporal evolution of burst rate }    \label{sec:activity_variation}
For each epoch, we measure the burst rate as the number of bursts detected per hour with fluence above the 90\% completeness threshold. Since the available number of uGMRT antennas varies from epoch to epoch, we adjust the fluence threshold accordingly. Specifically, we scale the fluence threshold $F_{\rm thres,24nov}$ (from 2022 November 24, the epoch used to estimate the fluence thresholds) by the ratio of antennas in each epoch given by
\[
F_{\rm thres,\,epoch} \;=\; F_{\rm thres,24nov}\,\times\frac{N_{24\rm nov}}{N_{\rm epoch}}\;,
\]
where $N_{\rm epoch}$ is the number of antennas used in that epoch and $N_{24\rm nov}$ is the number used on 2022 November 24. We then count the number of bursts above $F_{\rm thres,\,epoch}$ and divide by the on‑source time to obtain the burst rate. This procedure is carried out separately for band-3 and band-4.

Furthermore, the fluence threshold of 2.5 ms is used for bursts with width below 2.5 ms, that of 5 ms is used for bursts with widths between 2.5 ms and 5 ms, that of 7.5 ms used for bursts with widths between 5 ms and 7.5 ms, and that of 10 ms is used for bursts with widths greater than 7.5 ms. We adopt this approach because the completeness threshold increases with burst width, i.e., the threshold at a given width is greater than the thresholds at all smaller widths. This approach ensures a consistent and conservative selection across different width ranges, maintaining the intended completeness of the sample.

\par
\begin{figure*}
    \centering
    \includegraphics[width=\textwidth]{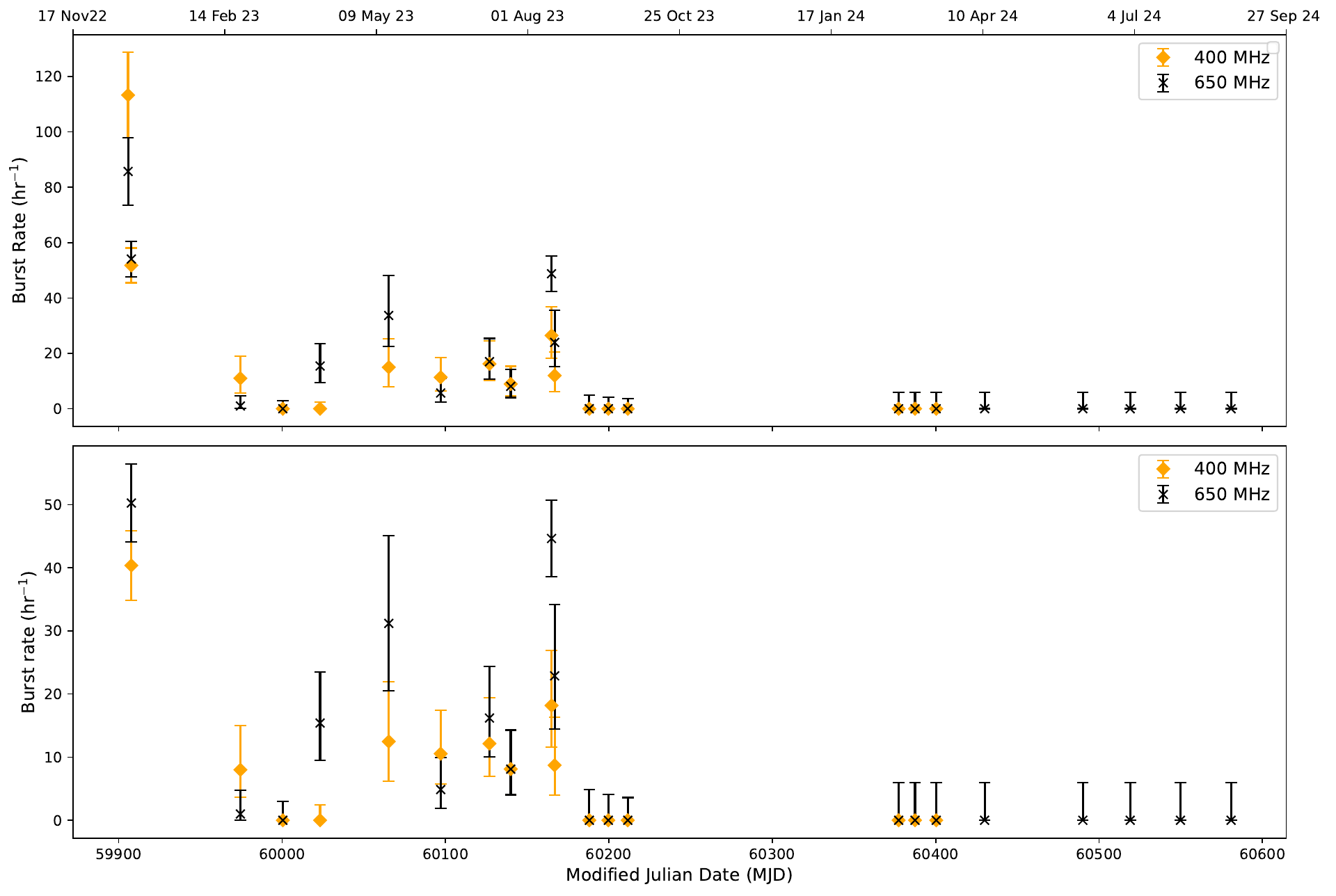}    
    \caption{ Burst rate as a function of time during the observation campaign from 2022 November 22 to 2024 September 27 is shown for both band-3 (300-500 MHz) and band-4 (550-750 MHz). Top: the burst rate calculated based on all the bursts detected above the respective 90\% completeness thresholds for different widths, as explained in the text. Bottom: The burst rate at each epoch is calculated based on the highest fluence threshold among all widths and both the bands, to provide a meaningful comparison of the rates at the two bands. The single band observations at band-3 and band-4 on 2022 November 22 are excluded. }
    \label{fig:activity_variation}
\end{figure*}

Figure~\ref{fig:activity_variation} shows the temporal variation of the burst rate for FRB\,20220912A in band-3 and band-4 over the period from MJD 59905 to MJD 60581. During this period, we observe periods of high activity followed by low activity, indicating modulation of the activity level. The observed activity pattern is similar to that of known active repeaters such as FRB\,20180916B and FRB\,20121102A \citep{Li_2021_periodicitymodel,Barkov_2022_periodic_model}. Notably, a significant increase in burst rate is seen around MJD 60166, comparable to the high activity observed near MJD 59905. However, following MJD 60188, the burst rate drops to near-zero levels and remains at that level for an extended duration till our latest observation on MJD 60581. \citet{abbott2026radiomonitoringcampaignactive} report that the source remains active at least till 60410. While activity was sustained, the final phase after 60225 was characterized by a very low burst rate consistent with our observation, suggesting a potential transition into an almost quiescent state. According to the detections from our observations and those reported by \citet{abbott2026radiomonitoringcampaignactive}, it appears that FRB\,20220912A was in an active state for more than 1.5 years before entering a quiescent or very low active phase. However, regular monitoring is crucial to see if the source becomes active again.

\par
\begin{figure*}
    \centering
    \includegraphics[width=0.9\textwidth]{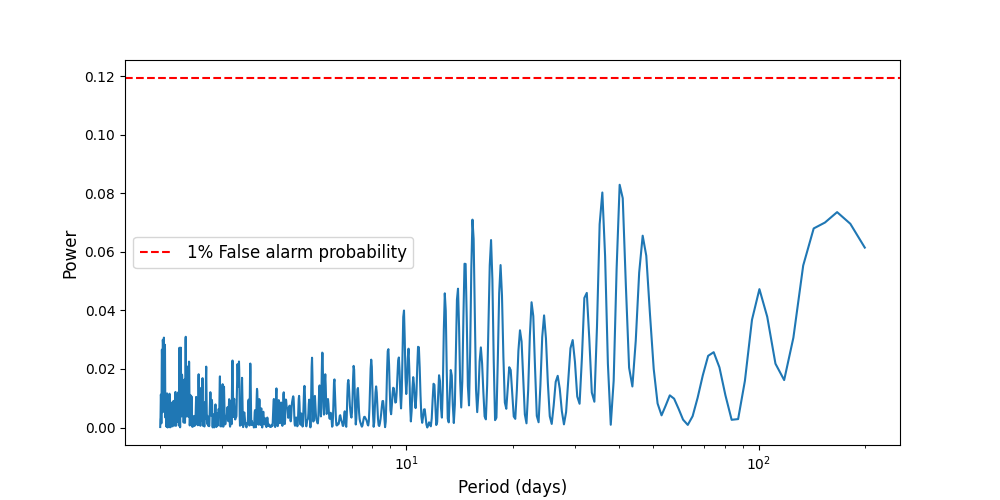}    
    \caption{Lomb Scargle periodogram obtained using the number of bursts detected by CHIME/FRB in each day from FRB\,20220912A during the period from MJD 59833 to MJD 60269. The red dashed line indicates the 1\% false alarm probability.}
    \label{fig:period_search_with_CHIME}
\end{figure*}

The observed modulation in burst activity might suggest an underlying periodicity, but our varying observation cadence and number of observations make it difficult to detect such periodicity. To investigate this further, we analysed publicly available data for FRB\,20220912A from CHIME/FRB\footnote{\url{https://www.chime-frb.ca/repeaters}}, spanning MJD 59833 to 60269. We created a time series of the number of bursts detected per day, and then obtained a Lomb-Scargle periodogram using the astropy \texttt{LombScargle} package. Figure~\ref{fig:period_search_with_CHIME} shows the resulting periodogram, which reveals no statistically significant periodicity in the activity level. We note that the information regarding the instrument's downtime is not publicly available, and hence, not folded in the above analysis.



\subsection{ Energy distribution }  \label{sec:energetics} 

The energy distribution of bursts from repeating FRBs offers critical insights into their origins and emission mechanisms. The energy distributions of certain repeating FRBs closely resemble the energy distribution obtained for non-repeating ones. This similarity implies a potential link between these populations \citep{kirsten2023connecting}. In several cases, the low activity level of the FRB results in the detection of statistically insignificant bursts, making it difficult to study the temporal evolution of the energy distribution \citep{sand2023chimefrbstudyburstrate,bhattacharyya2024widebandmonitoringfrb180916j01586}. However, extreme repeaters like FRB\,20220912A remain in high activity for a longer duration. This extended activity helps in robustly estimating the distribution's model parameters for each epoch and their variation with time.
\par
To determine the cumulative distribution, we consider the bursts above the 90\% completeness threshold for 2.5 ms, 5 ms, 7.5 ms and 10 ms for each epoch as explained in the previous subsection. The number of bins is chosen to be equal to the square root of the total count. Then, we utilize the \texttt{scipy. optimise. minimise} function to fit the cumulative distribution of fluences for bursts using a single power law or a broken power law, incorporating Poisson errors as outlined in \citet{cash_stat}, with square root uncertainty for numbers greater than 30. We choose the best model between these two using Akaike’s information corrected criterion \citep[AICC;][]{PORTET2020111}, similar to that done in \citet{lal2025lowenergyradioburstsmagnetar}. AICC considers the goodness of fit, the complexity of the model (number of fitted parameters), and the number of data points to be fitted to help in choosing the best model.

After selecting the best model, the model parameters are characterized using Markov Chain Monte Carlo (MCMC) sampling. The MCMC approach allows incorporating asymmetric uncertainties for pulse counts below 30 and the square root of the counts for larger numbers. For the MCMC parameter estimation, we have used the \texttt{emcee} package \citep{Foreman-Mackey_2013} with the best-fit estimates from the \texttt{scipy.optimize.minimize} as the initial guesses for the chosen best model. The asymmetric uncertainties on the data points are used while defining the likelihood functions. The MCMC estimation gives a posterior distribution for the model parameters, and the respective medians are chosen as the best-estimated parameters, along with the associated error bars deduced from the 84 and 16 percentiles. The posterior distributions for the model parameters and any covariances between them were examined manually for each of the observations, and the fitted parameters were found to be well constrained for all the epochs. 

\begin{figure*}
    \centering
    \includegraphics[width=0.45\textwidth]{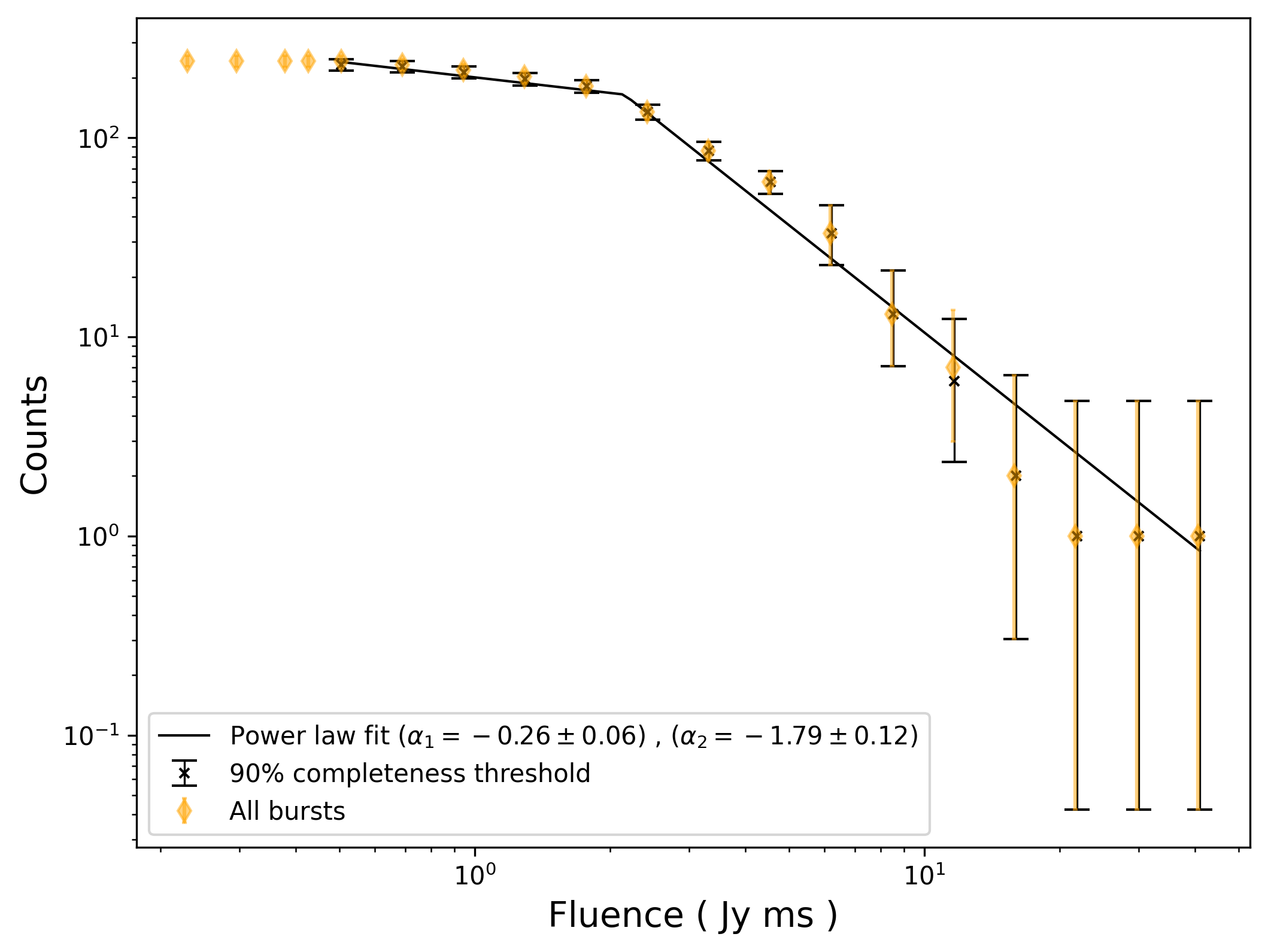}
    \hfill
    \includegraphics[width=0.45\textwidth]{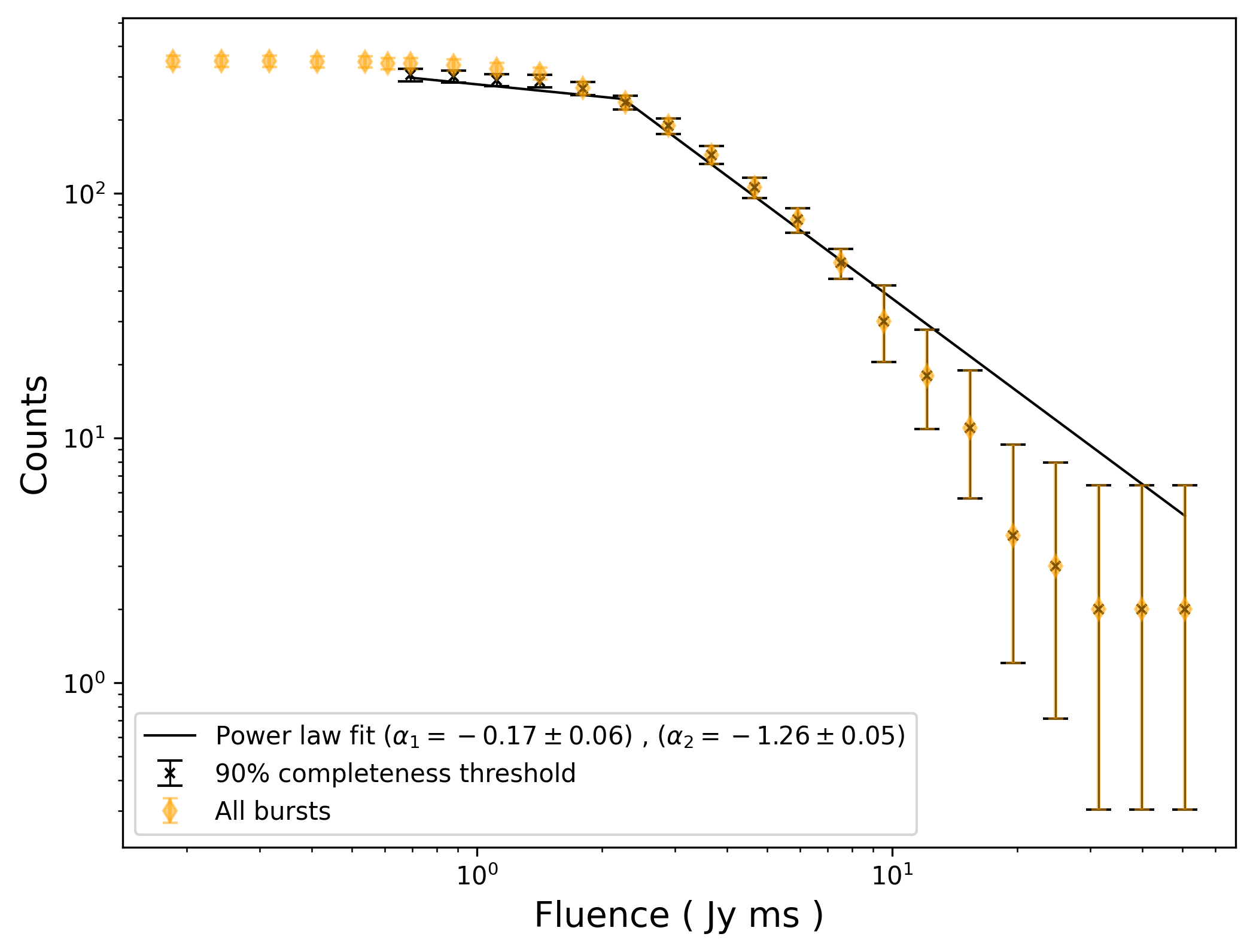}
    \caption{Cumulative distribution of fluences for all the bursts above the completeness threshold detected during the observing campaign at band-3 (\textit{left}) and band-4 (\textit{right}) are shown as black points. Black lines indicate the fitted broken power law. The yellow points indicate the cumulative distribution of fluences for all the detected bursts.}
    \label{fig:energydistribution_allepochs}
\end{figure*}

For most epochs, the cumulative distribution fits well with a single power law. However, for MJD 59907 at band-3 and MJD 60164 at band-4 broken power law is a better fit. The cumulative distributions of fluences for all the bursts over the whole observing campaign above the 90\% completeness threshold is well-fitted with a broken power law with two exponents of -0.26 $\pm$ 0.06 and -1.79 $\pm$ 0.12 for band-3, and -0.17 $\pm$ 0.06 and -1.26 $\pm$ 0.05 for band-4, as shown in Figure~\ref{fig:energydistribution_allepochs}. The break in the distribution occurs at 2.9 $\times$ $10^{-29}$ ergs Hz$^{-1}$ and 3.1 $\times$ $10^{-29}$ ergs Hz$^{-1}$ for band-3 and band-4 respectively. The single band observations on MJD 59905 in band-3 is well-fitted by a single power law with slope -1.37 $\pm$ 0.1 and in band-4 with slope -0.9 $\pm$ 0.1. In our observational campaign, there were four other epochs where we could fit a single or broken power law to the cumulative distribution of fluences at both band-3 and band-4 and two more, but only in band-4. In Appendix~\ref{sec:plots_for_ind_epochs}, we provide the cumulative fluence distributions for each epoch and the corresponding power-law indices, along with an example corner plot demonstrating the well-constrained fitted parameters and their uncertainty regions.


\subsection{Second scale periodicity search}   \label{sec:periodicity}

For the repeating FRBs, the leading model for the progenitor is a neutron star with a high magnetic field, with the emission being coherent \citep{2017MNRAS.468.2726K}. Neutron stars, especially pulsars, are associated with beamed emission, with the emission cutting the line of sight periodically. We search for any underlying periodicity from FRB\,20220912A with arrival times of the bursts detected at band-3 and band-4 individually for each epoch. We also search for periodic emission where the individual single pulses could be buried under the noise, as seen in the case of faint pulsars and magnetars.  

\subsubsection{FFT search using time series}

We search for the faint periodic signal by de-dispersing the filterbank file for each session in band-3 and band-4 at a DM value of 220\,pc\,cm$^{-3}$. The de-dispersion is done on the filterbank file using subroutine {\tt prepdata} from {\tt PRESTO}. We use the {\tt PRESTO} standard routine {\tt acceelsearch} to search for any periodic signals without any acceleration. Consequently, the progenitor is implicitly assumed to be isolated or in a binary system with an orbital period much larger than the observation duration so that the observed period would not appear to change within the observation. We did not detect any genuine periodic signal with a significance greater than 3$\sigma$.

\subsubsection{Period search with burst arrival times} \label{sec:lombscargle_for_r117}
\begin{figure*}
    \centering
    \includegraphics[width=0.61\textwidth]{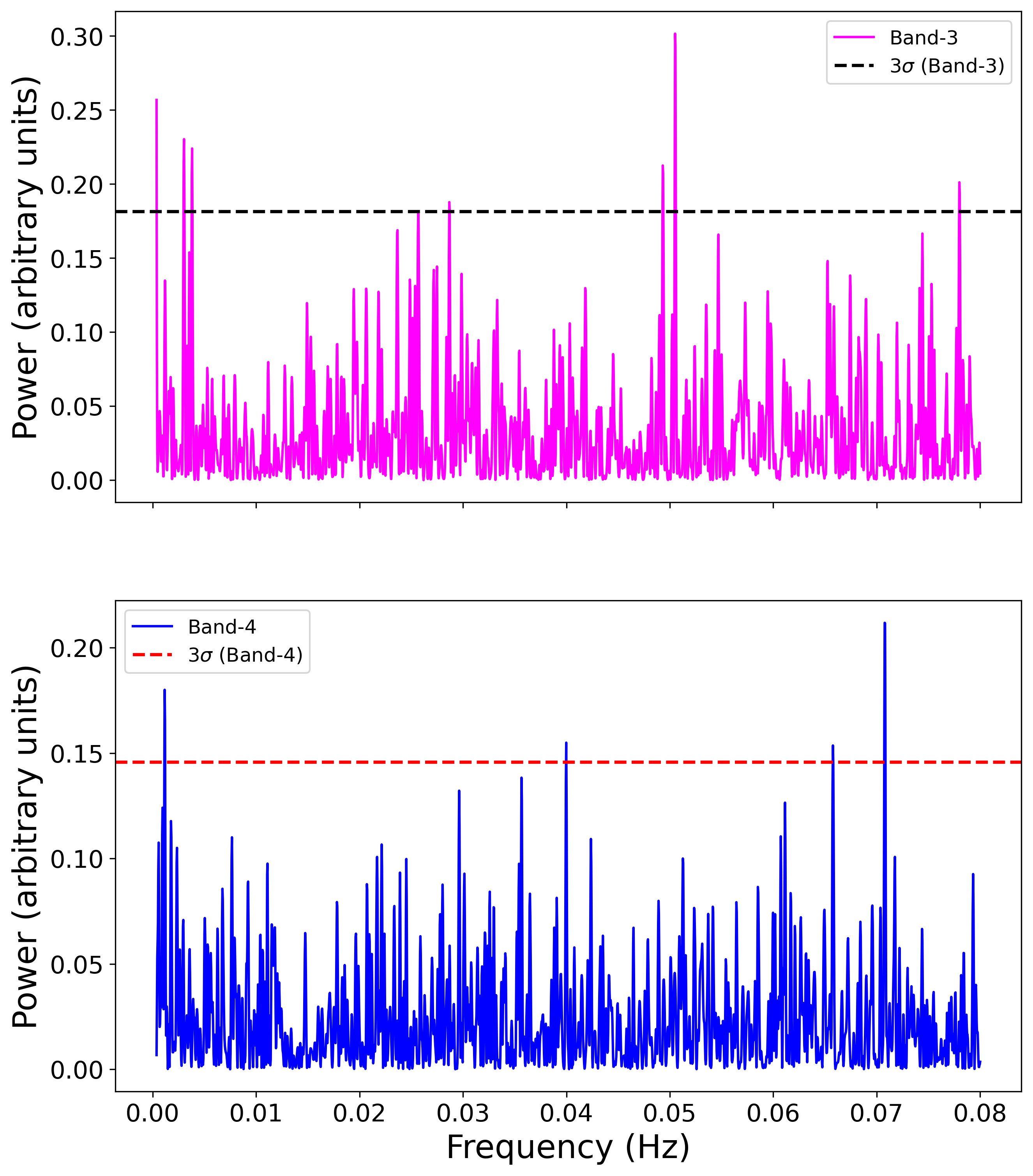}
    \vspace{0.2cm}
    \includegraphics[width=0.33\textwidth]{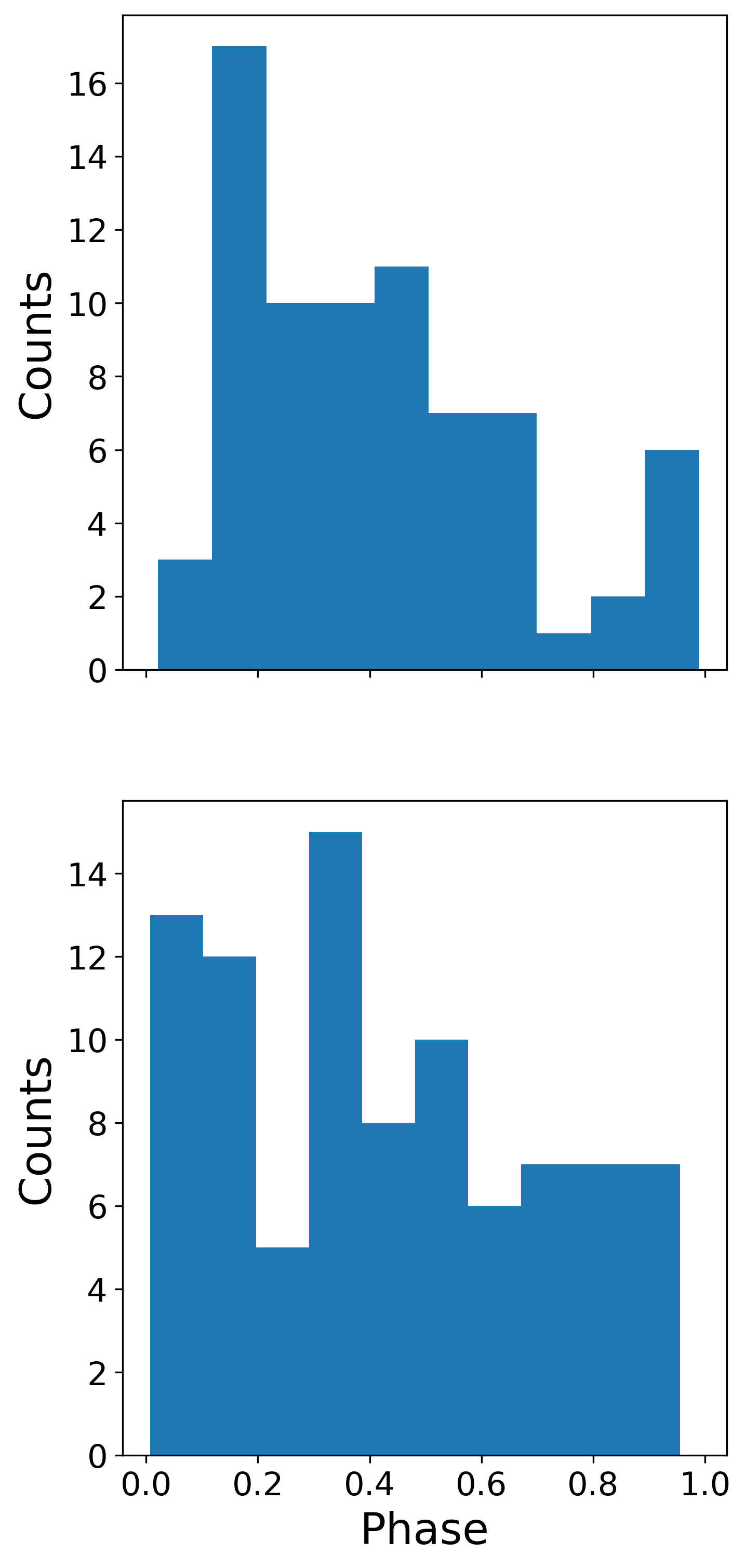}
    \caption{ Lomb Scargle periodograms obtained from the arrival times of the bursts detected on 2022 November 24 at band-3 \textit{(top left)} and band-4 \textit{(bottom left)}. The dashed horizontal lines in these periodograms indicate the estimated 3$\sigma$ threshold. \textit{Top right:} Histogram for the phases obtained corresponding to the brightest peak at 19.8\,s at band-3. \textit{Bottom right:} Histogram for the phases obtained corresponding to the brightest peak at 14.4\,s at band-4.}
    \label{fig:lombscargle}
\end{figure*}

The Lomb-Scargle method is best suited for searching periods in unevenly sampled data \citep{Lomb_1976,Scargle_1989}. For each individual epoch, we use the arrival times of each burst to estimate the Lomb-Scargle periodogram, giving equal significance to each burst. For the periodograms at each epoch, use bootstrapping to estimate the significance of individual points \citep{lal2025lowenergyradioburstsmagnetar}. We obtain 2000 realisations of the periodogram from randomly generated arrival times within the respective observation durations. We then determine the 3$\sigma$ confidence level as the median value of those at all frequency bins in the periodogram. This 3$\sigma$ confidence level is shown in left panels of Figure~\ref{fig:lombscargle} as dashed lines. On 2022 November 24, we get peaks in the periodogram at more than 3$\sigma$ significance in both the bands (see the left panels of Figure~\ref{fig:lombscargle}). The highest peaks occur around 0.05 and 0.07 Hz, corresponding to periods of 19.8\,s and 14.4\,s, at band-3 and band-4, respectively. We show the phase histogram for the 14.4\,s period at band-4, and we do not see any significant clustering of phases (see bottom right panel in Figure~\ref{fig:lombscargle}). Similarly, when we obtain the histogram for 18.8\,s period, we see a broad clustering of the phases spread over the range from 0.1 to 0.6, indicating a possible periodicity with a large duty cycle. However, to establish if the periodicity is indeed true, particularly because this detection is at a very low significance, we need to detect the same period across multiple epochs or at different frequency band. However, we do not see similar periodicities (or other periodicities at high significances) at other epochs and bands. We also perform a similar search to detect any periodicity with arrival times of bursts that were detected from FAST with publicly available data \citep{Zhang2023r117fast}. We perform this search for each of their observing sessions. However, we do not find any periodicity above 5 sigma significance in any of their observing sessions. 

\section{Discussion}  \label{sec:discussion_r117}

FRB\,20220912A is an exceptionally active repeating FRB that exhibited high levels of activity for several months since its discovery on 2022 September 12 by CHIME \citep{2022ATel15679....1M}. Our observations with uGMRT from 2022 November 22 to 2023 September 24 reveal modulation in activity, alternating between high and low phases (see Figure~\ref{fig:activity_variation}). After a five-month gap, we resumed observations from 2024 March 8 to 2024 August 28 with a monthly cadence but detected no further bursts, indicating that the source had entered a prolonged quiescent phase or at least it was not emitting bursts above our detection threshold (completeness thresholds roughly in the range 0.3 to 1.0 Jy\,ms; see Figure~\ref{fig:compthresholds}). The last reported detection by CHIME/FRB was in 2024 January, further supporting this inactivity above their fluence threshold ($\sim$\,5\,Jy\,ms). However, detections with CHIME/Pulsar indicate that the source had remained sporadically active but with a very low mean repetition rate, which also explains our non-detections in 2024.

\subsection{Energy Distributions}  \label{sec:energy_distributions_r117_discussion}
Several studies of active repeating FRBs indicate that their burst energy distributions are not well described by a single power law but instead require a broken power law, suggesting the presence of distinct emission regimes or mechanisms \citep{Li_2021, kirsten2023connecting, omar2024_r117}. However, interpreting these distributions requires caution due to various observational biases, such as telescope sensitivity, observing time, frequency coverage, observing bandwidth and completeness \citep{Wang_2019, 2021ApJ...920L..18A}. In our dataset, we measure a break at the lower end of the cumulative energy distribution (see Figure~\ref{fig:energydistribution_allepochs}), consistent with the findings from FAST for sources like FRB\,20220912A, FRB\,20121102A, and FRB\,20201124A \citep{Zhang2023r117fast, Li_2021}. If such a break is intrinsic to the source, then there are likely two distinct emission mechanisms for bursts with higher and lower energies. 
\par
\begin{figure*}
    \centering
    \includegraphics[width=0.8\textwidth]{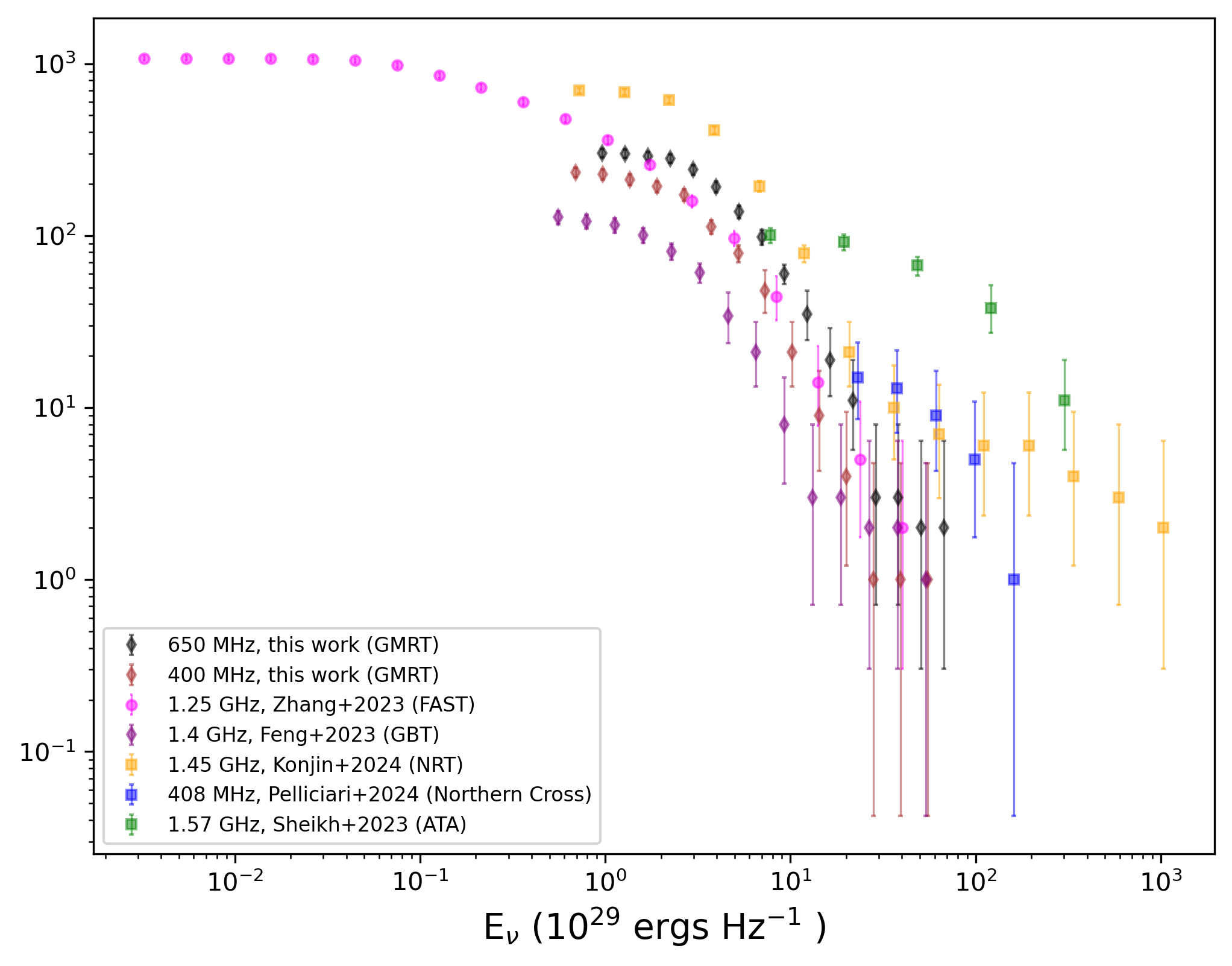}
    \caption{Comparison of the cumulative distribution of the spectral energies for FRB\,20220912A at different observing frequencies of 1.25 GHz \citep[magenta circles;][FAST]{Zhang2023r117fast}, 1.4 GHz \citep[brown diamond;][GBT]{Feng2023_r117gbt}, 1.45 GHz \citep[yellow squares;][NRT]{konjin2024_nrt_r117}, 408 MHz \citep[purple squares;][Northern Cross Radio Telescope]{pelliciari2024_northencross}, 1.57 GHz \citep[green squares;][ATA]{Sheik2024r117Allen}, and at 400\,MHz and 650\,MHz from this work (GMRT).}
    \label{fig:energydistribution_compare_with_other_telescopes}
\end{figure*}

In Figure~\ref{fig:energydistribution_compare_with_other_telescopes}, we compare the cumulative spectral energy distributions of bursts detected at 400 MHz (band-3) and 650 MHz (band-4) from FRB\,20220912A with those reported at different frequencies using other telescopes. The isotropic spectral energy, $E_{\nu}$, of a burst is calculated from the
observed fluence following
\begin{equation}
E_{\nu} = \frac{4\pi D_L^2 F_\nu}{(1+z)} ,
\end{equation}
where $E_\nu$ is in ergs\,s, F is the burst fluence in Jy\,ms, $D_L$ is the luminosity distance in cm, $z$ is the redshift and $\nu$ is the centre frequency of the band. The overall distribution matches well with those from highly sensitive instruments like FAST and GBT, while less sensitive single-dish telescopes such as NRT report a break at higher spectral energies, close to the upper energy limit of our dataset and FAST \citep{omar2024_r117}. 
\par
The data compiled in Figure~\ref{fig:energydistribution_compare_with_other_telescopes} comprises of observations at different frequencies and times and using instruments with different sensitivities. The activity level of the source varies with time, and that gives rise to vertical offsets. Accounting for these vertical offsets, we note that the cumulative distributions across a wide frequency range and different observing times have broadly similar shapes. Moreover, despite the wide frequency range and different observing epochs, a break at around the spectral energies (1$-$2)$\times 10^{29}$\,erg\,Hz$^{-1}$ is also evident in nearly all the data sets. Hence, the break does not only appear to be intrinsic to the emission, it also stays consistent with time. 
\par
The power-law slopes on the two sides of the break in various observations are also similar, however, there are some notable variations too. Given very different observing setups and instruments used for multiple data sets, and the non-availability of information on completeness analysis for most of the cases, it is hard to quantify whether the variations are indeed intrinsic or just due to the above aspects. Nevertheless, a common break and the overall similarity of the distributions strongly suggest intrinsic emission characteristics that persist over at least several months, if not longer. There might be an analogy here with many of the Galactic neutron stars known to emit giant pulses along with the regular periodic pulses. The energy distributions of the giant pulses and the regular periodic pulses are often modeled as power law and lognormal distributions, and the distributions show broadly consistent characteristics across different frequencies and epochs \citep[power law indices for cumulative distributions in the range 1.3$-$2.3;][]{Mickaliger_2012}. However, energy distribution of the radio bursts from the Galactic magnetars is known to change rapidly from one epoch to another \citep[e.g.,][]{lal2025lowenergyradioburstsmagnetar}. But the Galactic magnetars have also been observed to emit giant pulse like emission only relatively infrequently. In the systematic study of the magnetar J1810$-$197, \citet{lal2025lowenergyradioburstsmagnetar} found the source to exhibit giant pulses at 15 different epochs (see their figure 4), and a power-law tail was required to fit the energy distributions at 8 of these epochs. At 5 of these epochs, where the energy distributions were well fitted by a broken power-law, the index at the tail end sampled a rather tight range (-1.85 to -2.1). Thus, it seems likely that the energy distribution characteristics of the giant pulse emission in magnetars might also remain consistent over time, despite the huge variability otherwise seen in their radio emission characteristics.
Overall, there could be a similarity in the consistency of energy distribution characteristics of giant-pulse emission from the Galactic neutron stars and the FRB~20220912A bursts. 

\par
\begin{figure}
    \centering
    \includegraphics[width=0.45\textwidth]{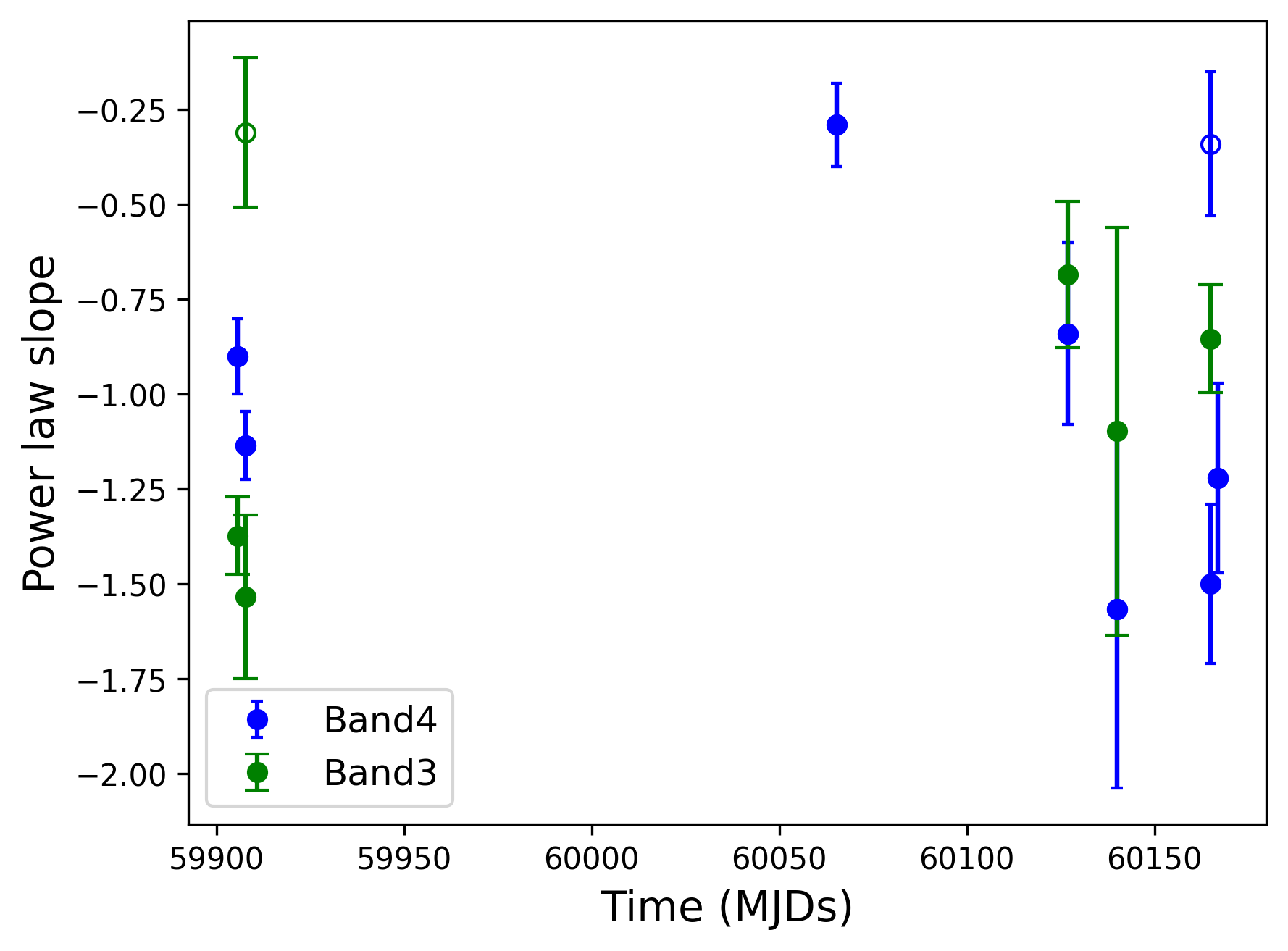}
    \caption{ Power law slope with time for both band-3 and band-4 during the period from MJD 59905 to MJD 60188. The hollow marker represents the power law slope for the lower energies when the distribution is well-fitted with a broken power law, and solid circles indicate the slope for higher energies. A solid circle also indicates the slope obtained when it is well-fitted with a single power law. }
    \label{fig:power_law_slope_with_time}
\end{figure}

Although the energy distribution appears to follow a single power law within most individual epochs, the slope varies over time and between frequency bands without a clear monotonic trend, as shown in Figure~\ref{fig:power_law_slope_with_time}. This behaviour resembles to other sources like FRB\,20121102A, FRB\,20240114A, and the magnetar XTE\,J1810–197 \citep{Lanman_2022, Sang_2023}. In Figure~\ref{fig:power_law_slope_with_time}, we observe steeper slopes at 400 MHz than at 650 MHz initially around MJD 59907, but the trend reverses later around MJD 60168. However, these variations could have contributions from just the low number statistics at several epochs. In addition, there could be stochastic variations in the emitted bursts and one or the other emission mechanism (corresponding to the two power law indices around the break) might dominate at different epochs.

\subsection{Temporal and spectral variation in activity}  \label{sec:temporal_activity_r117_discussion}
Figure~\ref{fig:activity_variation} shows the modulation in the activity level of FRB\,20220912A during the period from MJD 59905 to MJD 60211, with a wide range of measured burst rates in the 300 to 750 MHz frequency range. Similarly, several other active repeaters have shown a wide range of burst rates at different frequencies. Since late 2023, observations with CHIME/Pulsar suggest that the baseline activity has become significantly low \citep{abbott2026radiomonitoringcampaignactive}, which is also consistent with non-detections with our observations in 2024. The sustained high activity over an extended duration supports progenitor models involving a young magnetar \citep{Beloborodov_2017,Margalit_2018,Metzger2019}, while the subsequent prolonged inactivity may suggest that the energy reservoir of the FRB source has been exhausted or significantly depleted for FRB production. Alternatively, progenitor models involving a binary system with long orbital periods could explain the activity variation, where interaction between the neutron star and the companion star leads to FRB production \citep{Wang_Zhang_2022_r67_binary}. Continued monitoring will be essential to catch the source during another active phase and to further test these hypotheses.  
\par
FRB\,20180916B and FRB\,20121102A show periodic activity modulated with cycles of 16.35 days \citep{chimeR3} and $\sim$157 days \citep{Rajwade_2020}, respectively. In the case of FRB\,20121102A, identifying the periodicity was challenging, as it was not active in every predicted active window, requiring long-term, high-cadence monitoring \citep{Rajwade_2020}. From Figure~\ref{fig:period_search_with_CHIME}, we infer that any underlying periodicity with periods up to around a hundred days or so is unlikely for FRB\,20220912A, but longer periods can not be ruled out. In the future, long-term high cadence observations with high sensitivity will be crucial for detecting any underlying periodicity for any active repeating FRB source, particularly if its emission is sporadic, i.e., not active in every cycle or if it has a long intrinsic period ($\geq$ one year). Alternatively, the modulation could arise from a narrow emission beam that wanders within a cone or funnel, potentially due to the dynamics of a precessing disk, as suggested for X-ray binaries \citep{katz_2017}.

\par
For FRB\,20220912A, we do not see any strong evidence of frequency-dependent activity. However, there is a hint of difference in the mean activity at our two observing bands. The mean burst rate for band-3 and band-4, considering only the bursts with a common threshold (bottom panel of Figure~\ref{fig:energydistribution_compare_with_other_telescopes}), are $12.4^{+1.1}_{-1.1}$\,hr$^{-1}$ and $16.3^{+1.3}_{1.2}$\,hr$^{-1}$, respectively. This difference in the mean burst rate at 650\,MHz and 400\,MHz indicates that for FRB~20220912A, on average, emission is preferred at higher frequency. 

\subsection{Energy budget calculations} \label{sec:energy_budget_r117_discussion}
Assuming there is isotropic emission from the source, it is possible to get rough estimates of the total energy output from the FRB source. We calculate the total energy output of the source using a typical emission bandwidth of $\Delta \nu \sim$ 100 MHz for both band-3 and band-4 bursts with a median fluence of 2.7\,Jy\,ms for the period extending from MJD 59905 to MJD 60211. We compute the median fluence over all bursts detected in both bands with fluences above the 90\% completeness threshold.
\par
Based on the number of bursts detected during our observation campaign, we calculate the mean burst rate of 36 hr$^{-1}$. Considering this mean burst rate and the median fluence of 2.7 Jy ms, the total energy output for 306 days is roughly 9.6 × 10$^{42}$ ergs in the frequency range of  300 to 750 MHz. Strictly, it is only a lower limit on the energy output as accounting for the bursts emitted at other frequencies and those below our detection threshold would further increase the energy output. Our estimated total energy output from the source is about 0.012\% of the total magnetic dipolar energy in the magnetosphere of a typical magnetar (see appendix~\ref{sec:energy_budget} for details). Here, we naively assume the beaming fraction and the radio radiation efficiency cancel each other out. Thus, from our energy estimates, it is quite plausible that a magnetar with a surface magnetic field of $\sim10^{15}$\,G can explain the burst emission from FRB\,20220912A on much longer timescales.

\subsection{Rate comparisons with other telescopes and active repeating FRBs} \label{sec:cmparision_of_activity_with_others} 

Among active repeating FRBs, sustained periods of extreme activity (say, burst rates $\geq$\,100\,hr$^{-1}$) are rare. On MJD 59905, we recorded peak burst rates of 113.2 hr$^{-1}$ and 85.7 hr$^{-1}$ above the completeness thresholds in our single band-3 and band-4 observations, respectively. On the following day (MJD 59906), FAST measured a peak burst rate of about 200 hr$^{-1}$ \citep{Zhang2023r117fast}.

To compare our results with the FAST observations, we extrapolated our observed burst rates to the FAST completeness threshold of 0.06\,Jy\,ms (95\% completeness for widths $<$20 ms) using the logN-logS relation. Using the power-law index $\alpha$ derived from our uGMRT band-4 observations on MJD 59905, the extrapolated burst rate is 585\,hr$^{-1}$, which is much higher than the $\sim$200 hr$^{-1}$ observed by FAST. This discrepancy suggests that the power-law index could vary across different frequencies or at very short timescales, consistent with our findings discussed in detail in Section~\ref{sec:energy_budget_r117_discussion} (also see Figure~\ref{fig:power_law_slope_with_time}). Another likely scenario could involve an additional break at lower energies. For example, if we consider the power-law reported by \citet{Zhang2023r117fast} at lower energies, the extrapolated burst rate becomes $\sim$300\,hr$^{-1}$, reducing the discrepancy significantly.
\par
While several repeaters have exhibited high burst rates, particularly in FAST observations, our uGMRT observations represent the first detection of such high rates for this repeater at low radio frequencies. Previous uGMRT studies of FRB\,20180916B and FRB\,20201124A reported lower peak burst rates \citep[][without considering the different fluence thresholds]{marthi2020,marthi2022}. FAST also measured peak burst rates of 542\,hr$^{-1}$ for FRB\,20201124A and 122\,hr$^{-1}$ for FRB\,20121102A \citep{zhou2022,Li_2021}. FRB\,20220912A exhibited a high burst rate for the initial few months after its discovery \citep{abbott2026radiomonitoringcampaignactive,zhang2023}. The presence of activity for a longer duration is evident from Figure~\ref{fig:activity_variation} as well as the CHIME detections. \citet{abbott2026radiomonitoringcampaignactive} show that FRB\,20220912A was active for more than a year. Compared to the activity level and duration of other repeating FRBs in the publicly available CHIME/FRB repeater data (after taking into account the mean exposure time), FRB~20220912A clearly stands out in terms of remaining active for a longer duration, with brief episodes of heightened activity. Such repeaters with persistent activity over long durations are rare, with FRB\,20190520B being another example that has been observed to be active for about 4 years by exploiting FAST's high sensitivity. The prolonged activity of FRB~20220912A suggests a potentially highly energetic or dynamically active progenitor, such as a young, highly magnetised magnetar.

\subsection{Periodicity and progenitor models} \label{sec:periodicity_r117_discussion}

Our search for a short-timescale periodicity in FRB~20220912A bursts, using both Fourier analysis of the time series and Lomb-Scargle periodograms derived using burst arrival times, did not yield any significant detection. No significant detection of a short timescale periodicity has been made so far in any of the highly active repeating FRBs, even when they are most active \citep{Du_2024}. It is important to consider that the underlying periodicity might be present but masked by various factors, such as precession, orbital motion in a binary system, or variations in the emission beam. However, precession and orbital motion would manifest themselves as appropriately longer timescale periodicities, and there is no obvious evidence for a long timescale periodicity for FRB~20220912A. Another possibility to mask any underlying periodicity, as discussed by \citet{lal2025lowenergyradioburstsmagnetar}, is that the bursts sample a large range of rotation phases (typically more than 60-70\% of the rotation period).
\par
FRBs could also be produced through magnetic reconnection events in the magnetosphere \citep{Mahlmann_2022}. In such a case, the rotation period of the neutron star is difficult to detect since the emission is powered by rapid magnetic reconnection events in a current sheet of a magnetar wind, which are not restricted to happen around any specific phase of the neutron star. 

\section{Summary}    \label{sec:conclusion}

This study presents the first-ever detections of bursts from FRB\,20220912A with uGMRT in Band-3 (300–500 MHz) and Band-4 (550–750 MHz). FRB\,20220912A showed more than a year long active window following its discovery. Our observation campaign spanning 675 days in the frequency range of 300 to 750 MHz with roughly monthly cadence resulted in the following key findings.

\begin{enumerate}

    \item From nearly two years of monitoring of FRB\,20220912A, from MJD~59905 to MJD~60580, we detected 643 bursts, with peak burst rate at 113\,hr$^{-1}$. The source remained sporadically active for about a year following its discovery, with an initial episode of elevated burst rates lasting a few months. The activity showed clear modulation thereafter, with intermittent periods of high activity, followed by an apparent quiescence since late 2023. However, no significant periodicity was detected in the burst activity, with periods up to about 100 days. A strict periodicity at shorter timescales, i.e, up to several tens of seconds, was also not detected at any of the epochs.
    
    \item The cumulative energy distribution of bursts in both the bands revealed a break at lower energies, as has been seen earlier at different frequencies, and consistent with the trends observed in other repeaters like FRB\,20121102A and FRB\,20240114A. We also show that the cumulative energy distribution shape remains broadly consistent across different frequencies and over long timescales, even as the activity level varies significantly. However, the power-law indices characterizing the distributions measured at individual epochs can be different, likely due to limited burst counts or stochastically different combinations of the underlying emission mechanisms. 

    \item FRB\,20220912A remained active for an extended duration, exceeding 1.5 years, before entering a state of inactivity or emitting bursts at lower energies. Such prolonged activity for more than a year has been observed for the first time for a repeating FRB at lower frequencies ($<1$\,GHz). Based on our estimates of the total energy output of the FRB source, a magnetar with a surface magnetic field of 10$^{15}$G can easily power the radio burst emission from FRB\,20220912A. This suggests that a young magnetar is a likely progenitor for FRB\,20220912A. 
    
\end{enumerate}

\begin{acknowledgments}

\section*{Acknowledgments}
AK would like to thank Sujay Mate for useful discussions. We would like to thank the Centre Director and the observatory for the prompt time-allocation and scheduling of our observations. YM acknowledges support from the Department of Science and Technology via the Science and Engineering Research Board Startup Research Grant (SRG/2023/002657). We acknowledge the Department of Atomic Energy for funding support, under project 12$-$R\&D$-$TFR$-$5.02$-$0700. GMRT is run by the National Centre for Radio Astrophysics of the Tata Institute of Fundamental Research. 

\end{acknowledgments}

\section{Software and third party data repository citations} \label{sec:cite}
\facilities{Giant Meterwave Radio Telescope, Khodad, Pune }

\software{astropy \citep{2013A&A...558A..33A},  
          matplotlib \citep{Hunter:2007},
          scipy \citep{2020SciPy-NMeth},
          your \citep{Aggarwal2020},
          RFIClean \citep{Maan_2021},
          PRESTO \citep{2011ascl.soft07017R}
          }

\bibliography{bibliography}{}
\bibliographystyle{aasjournal}


\appendix

\section{Energy budget calculations} \label{sec:energy_budget}

For a perfect dipolar magnetic field, the energy density is given by, 
\begin{equation}
    U_{B} = \frac{ B^{2}}{8\pi} 
\end{equation}
here, B is the magnetic field. B is given by 
\begin{equation}
    \vec{B} = \frac{\mu}{r^{3}} ( 2\cos{\theta}\, \hat{r} + \sin{\theta}\,\hat{\theta})
\label{eq:magnetic_moment}
\end{equation}
here, $\mu$ is the magnetic moment, r is the radial distance from the centre of sphere and $\theta$ is the angle between the magnetic axis and the position vector. 

The total energy in the magnetosphere can be estimated as 
\begin{equation}
    E = \int U_{B}  dV  \approx \int_{R}^{inf} \frac{B^{2}}{8\pi}dV 
\end{equation}
After substituting B from equation\ref{eq:magnetic_moment} into the above equation, the total energy is equals to 
\begin{equation}
    E  \approx  \frac{\mu^{2}}{3R^{3}}
\end{equation}
In this equation $\mu$ can be replaced by B$_{s}$R$^{3}$/2, where B$_{s}$ is the surface magnetic field and R is the radius of the neutron star, so the total energy is equal to 
\begin{equation}
    E\approx\frac{B_{s}^{2}R^{3}}{12}
\end{equation}
For a magnetar, we can take the radius to be 10 km and the surface magnetic field of 10$^{15}$\,G; the total energy comes out to be 8.3 $\times 10^{46}$ erg.

\newpage
\section{Energy distributions, fits and completeness thresholds for individual epochs} \label{sec:plots_for_ind_epochs}

\begin{figure*}[ht]
    \centering
    \includegraphics[width=0.47\textwidth]{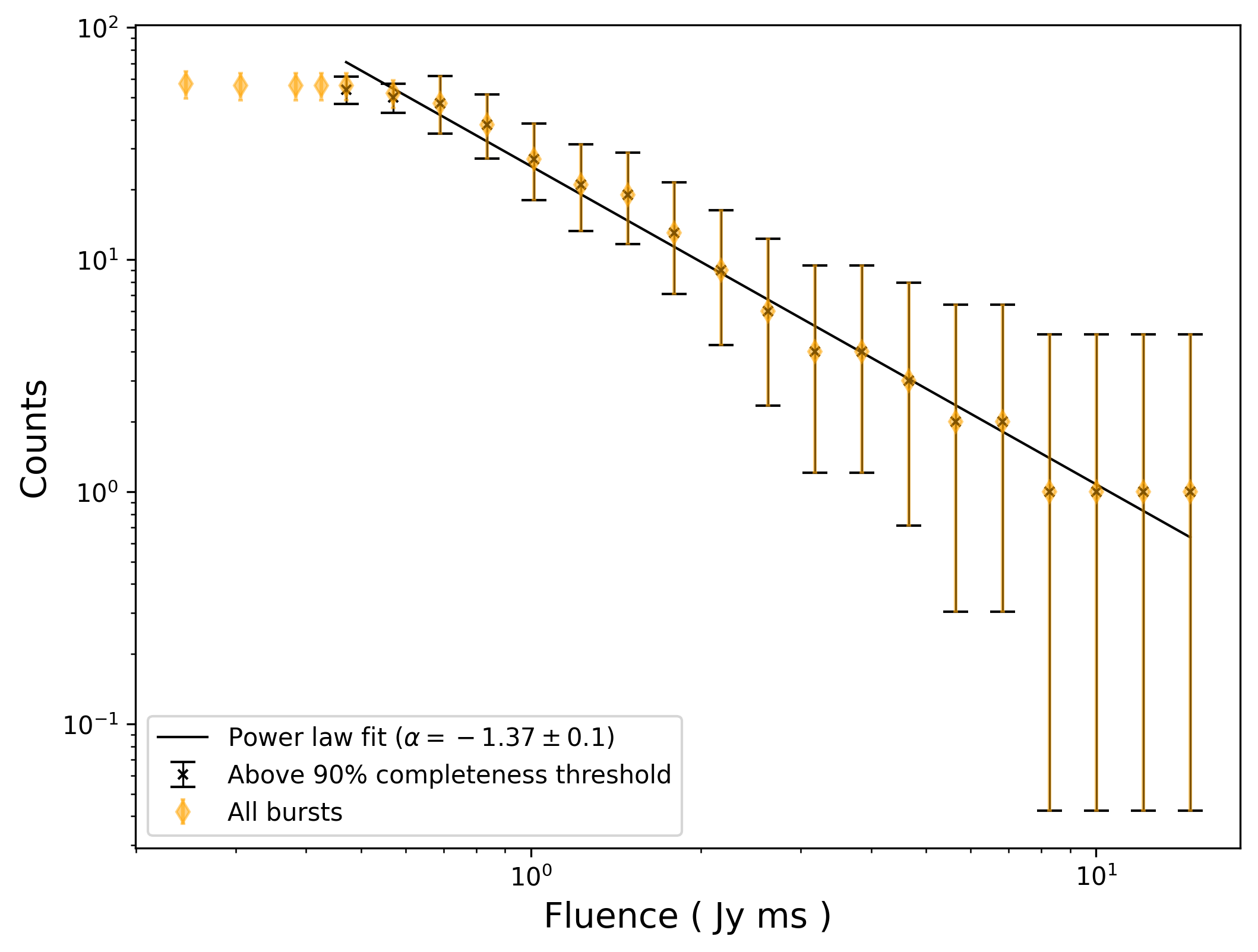}
    \hfill
    \includegraphics[width=0.47\textwidth]{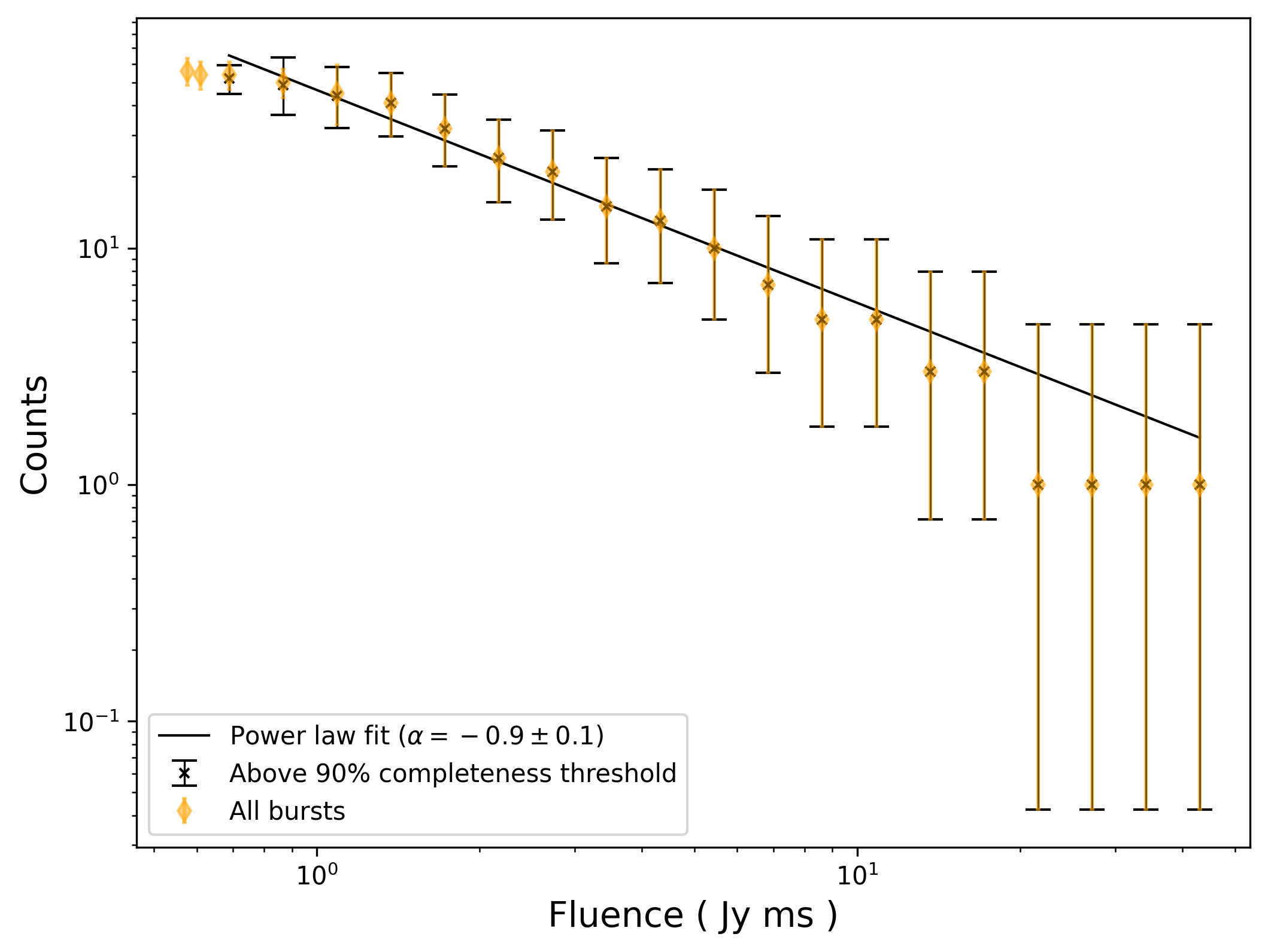}
    \caption{Cumulative distribution of fluences for all the bursts above the completeness threshold detected on 2022 November 22 (MJD 59905), for band-3 \textit{(left)} and band-4 \textit{(right)}. Black lines indicate the fitted model. The yellow points indicate the cumulative distribution for all the detected bursts, while the black points indicate that only for the bursts above the completeness thresholds.}
    \label{fig:energydistribution_singleband}  
\end{figure*}

\begin{figure*}{}
    \centering
    \includegraphics[width=0.47\textwidth]{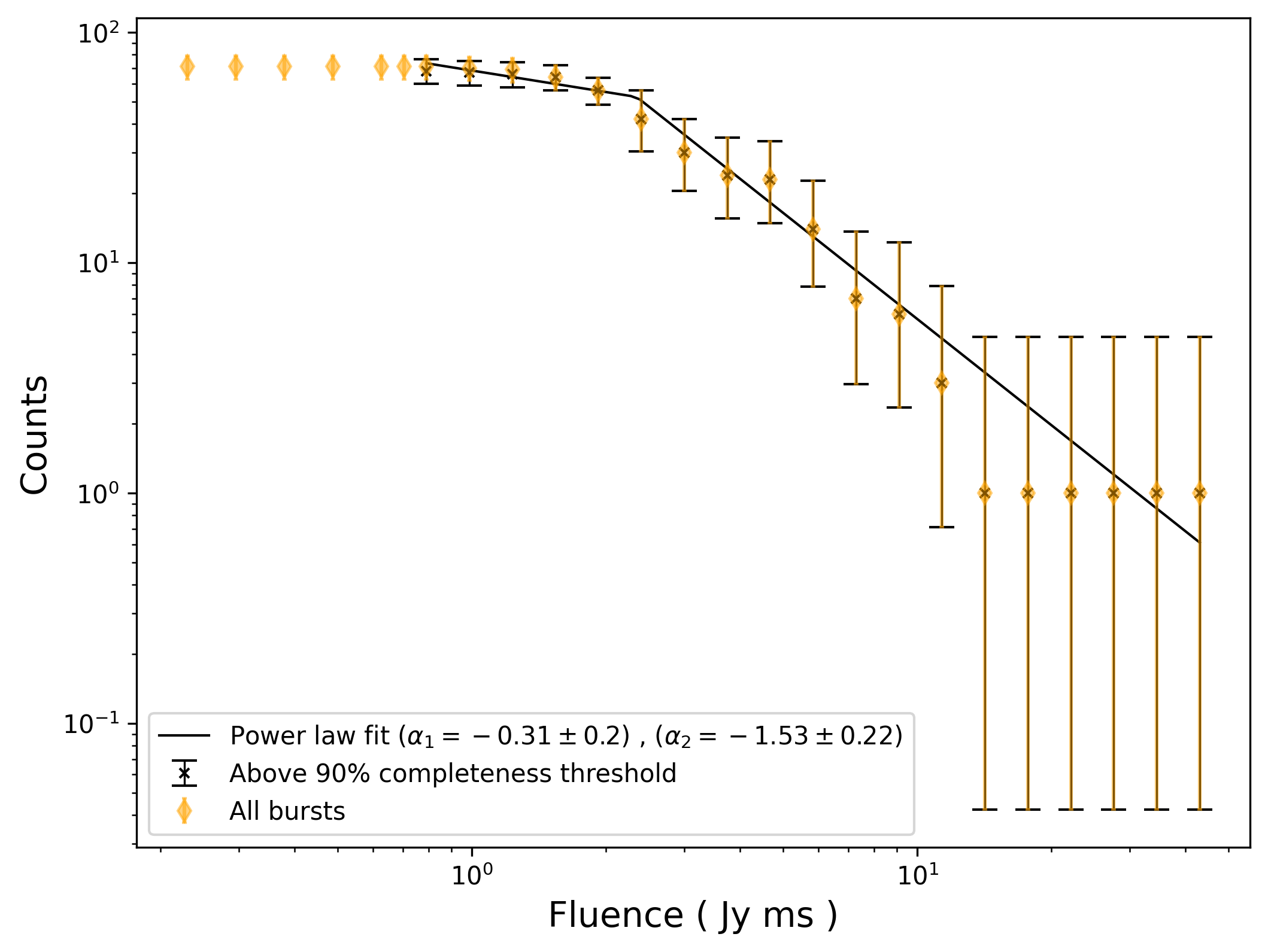}
    \hspace{0.2cm}
    \includegraphics[width=0.47\textwidth]{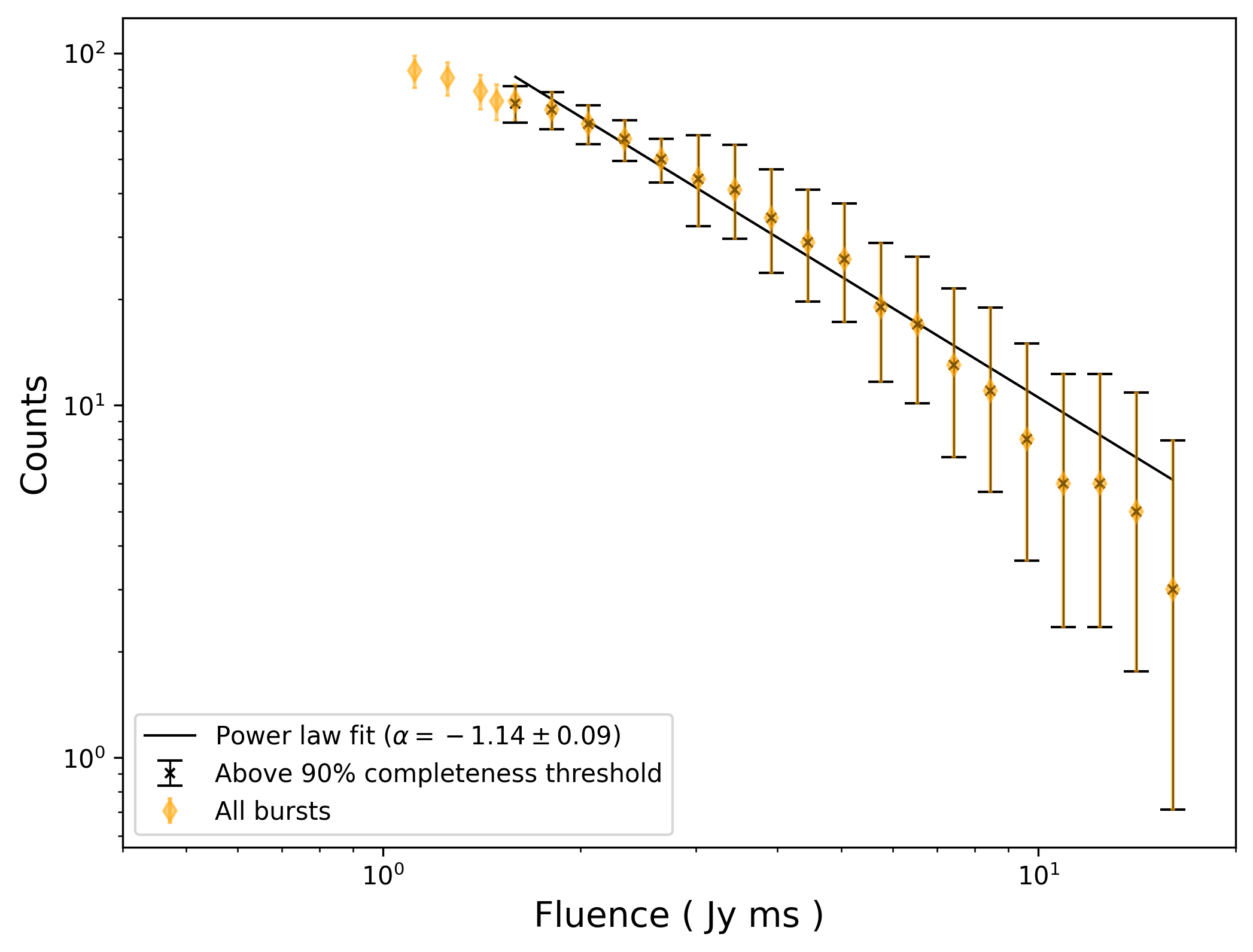}
    \caption{Same as Figure~\ref{fig:energydistribution_singleband} but for observations on 2022 November 24 (MJD 59907).} 
    \label{fig:energydistforindiepochs1} 
\end{figure*}

\begin{figure*}
    \centering
    \includegraphics[width=0.47\textwidth]{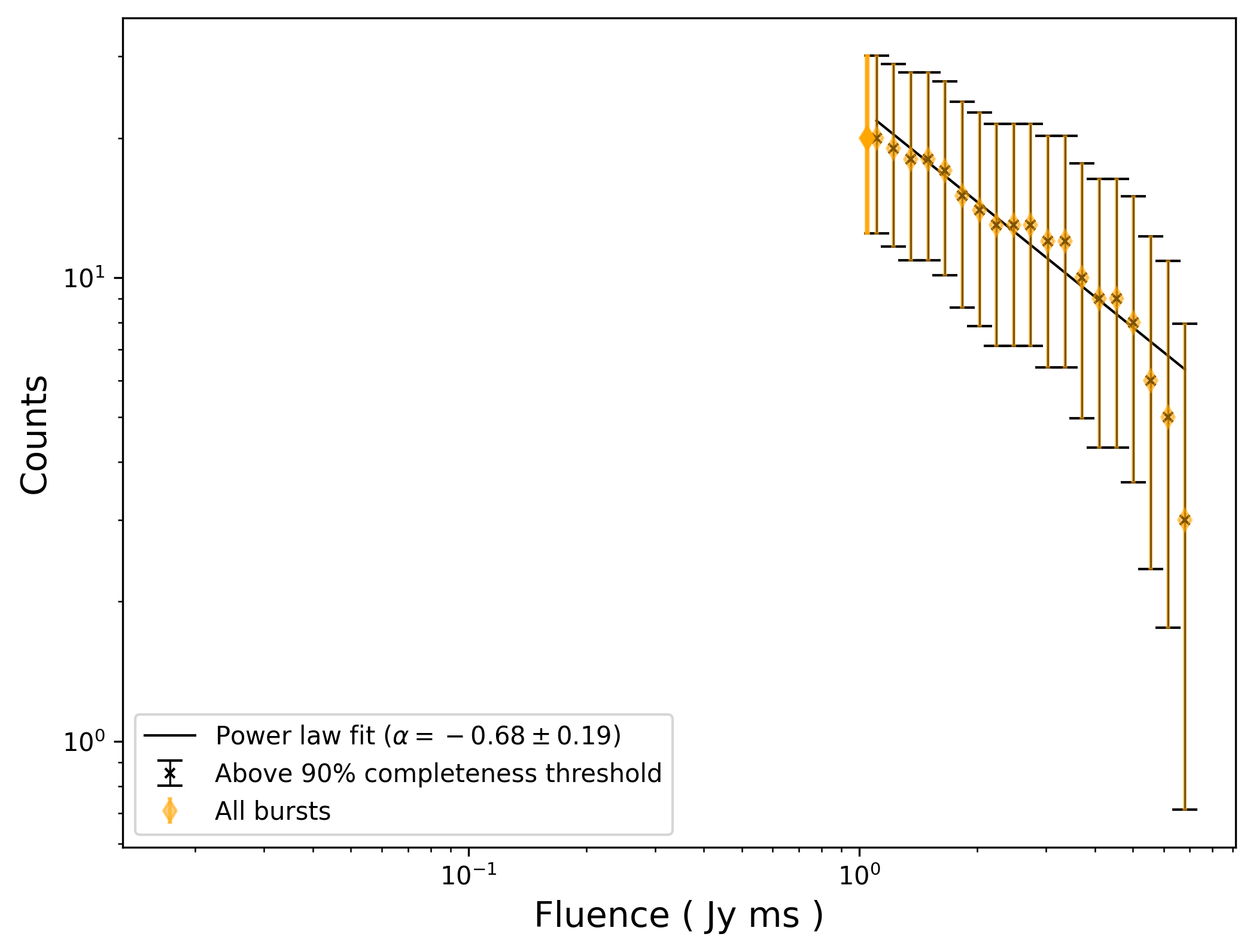}
    \hspace{0.2cm}
    \includegraphics[width=0.47\textwidth]{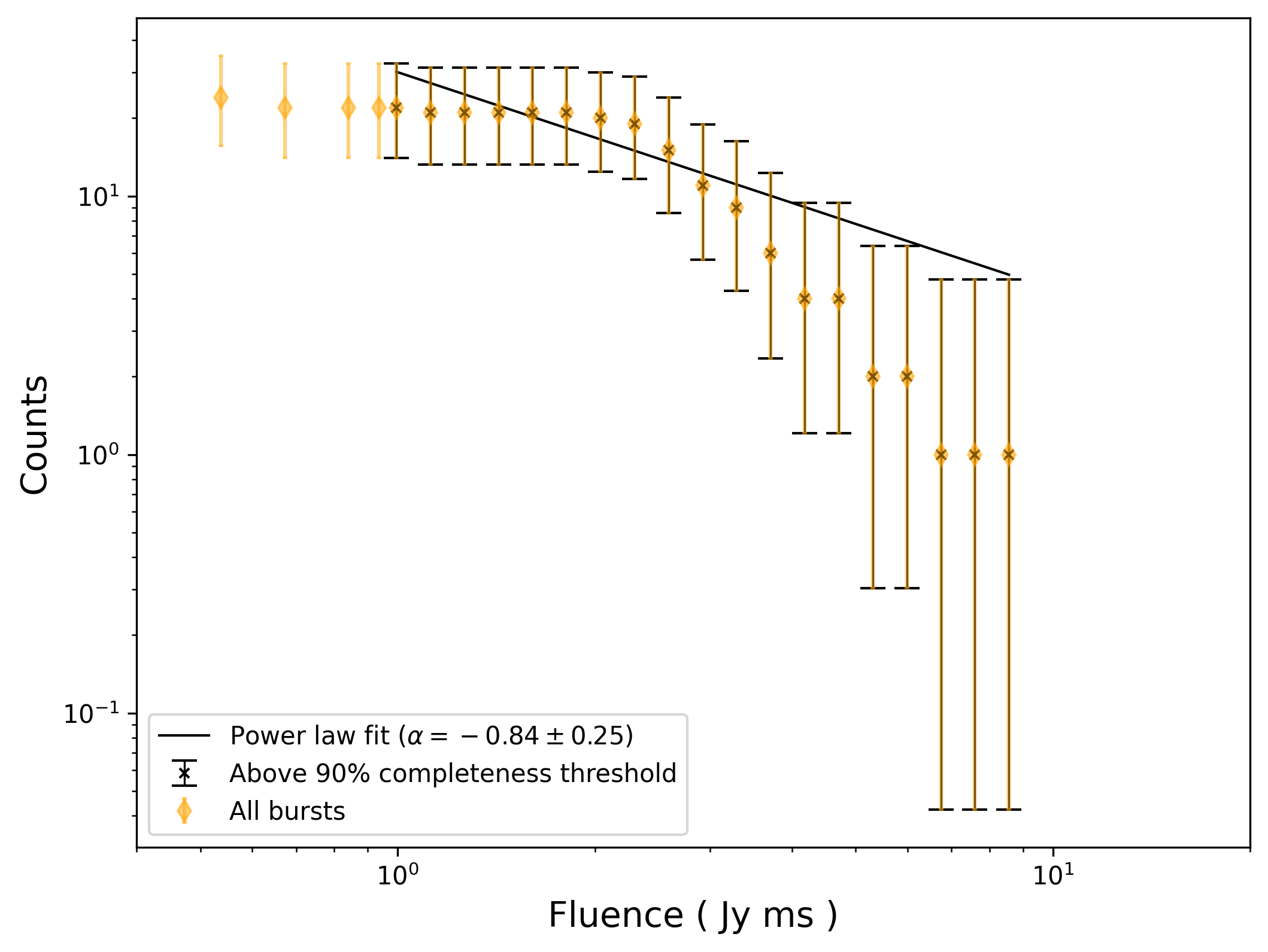}
    \caption{ Same as Figure~\ref{fig:energydistribution_singleband} but for observations on 2023 July 02 (MJD 60126). } 
    \label{fig:energydistforindiepochs2}
\end{figure*}

\begin{figure*}
    \centering
    \includegraphics[width=0.47\textwidth]{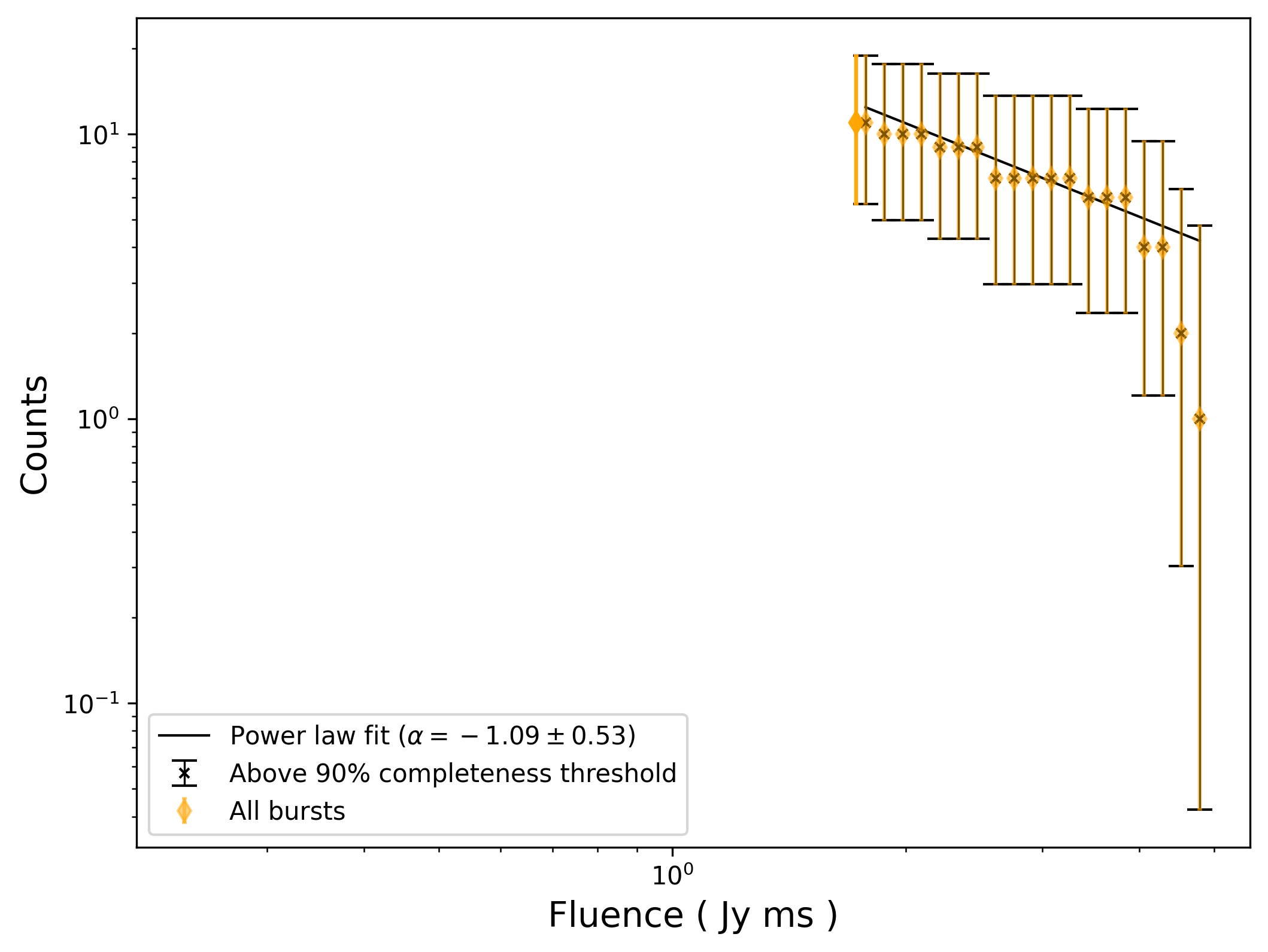}
    \hspace{0.2cm}
    \includegraphics[width=0.47\textwidth]{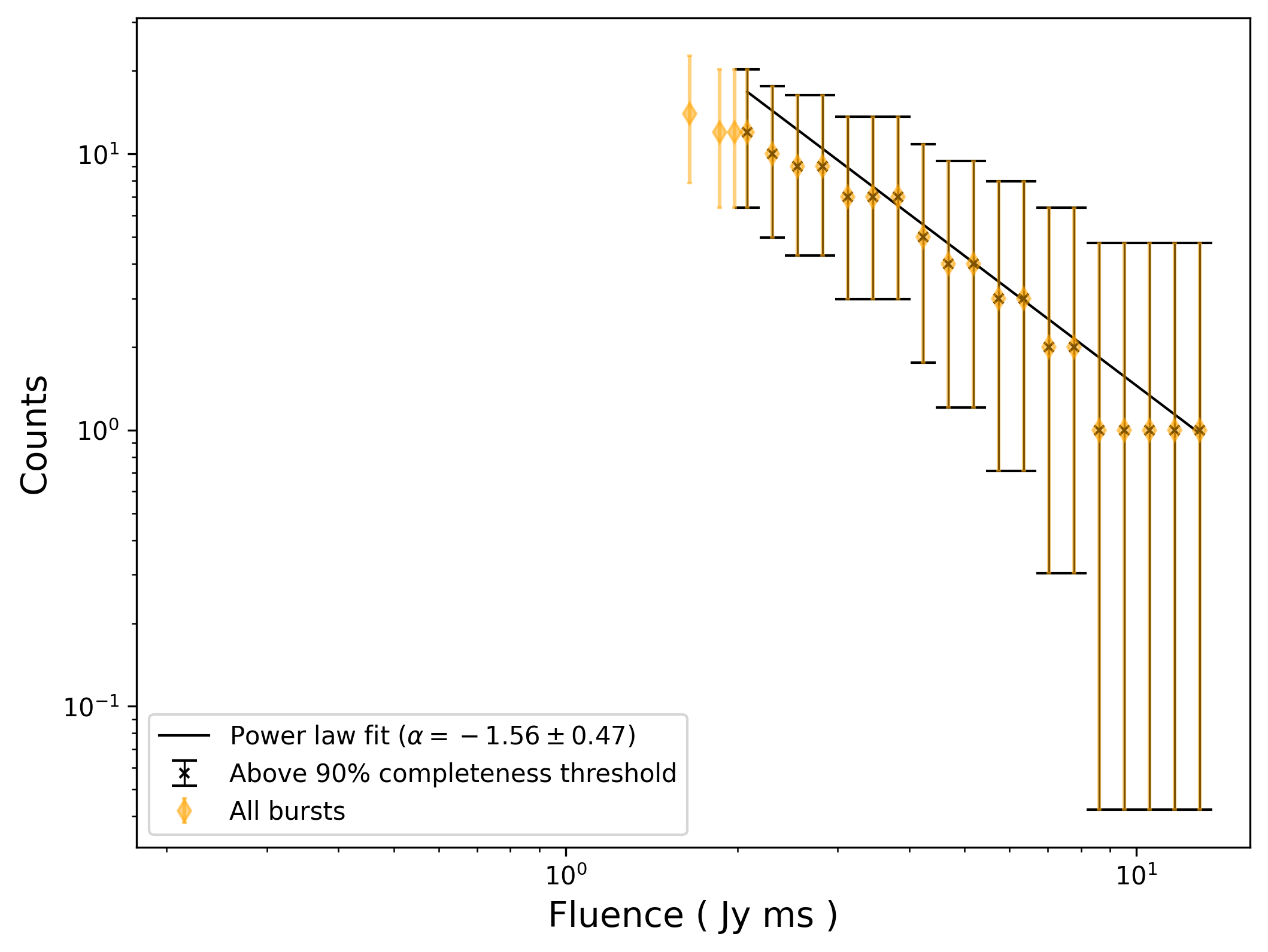}
    \caption{Same as Figure~\ref{fig:energydistribution_singleband} but for observations on 2023 July 15 (MJD 60139).} 
    \label{fig:energydistforindiepochs3}
\end{figure*}

\begin{figure*}
    \centering
    \includegraphics[width=0.45\textwidth]{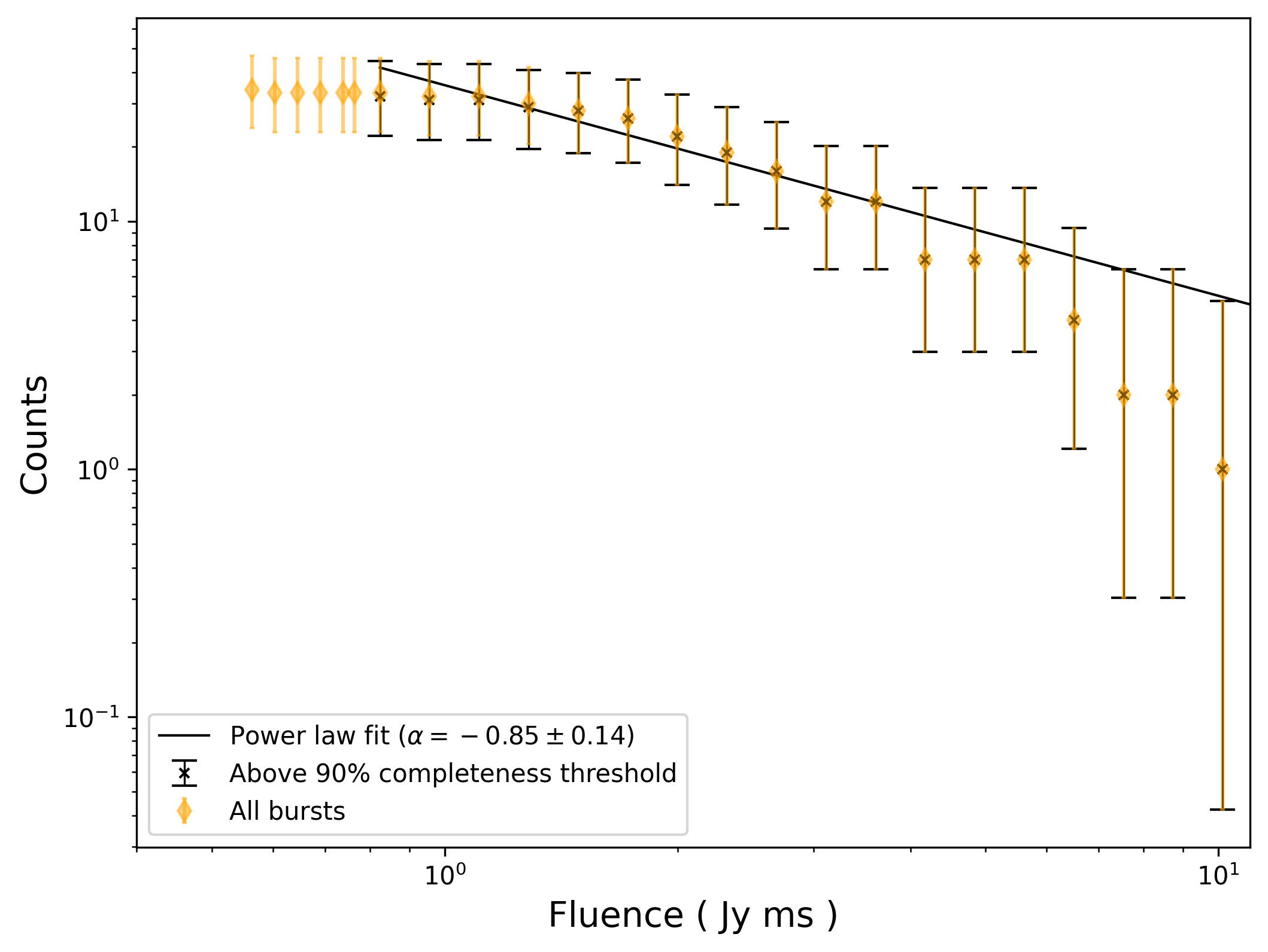}
    \hspace{0.2cm}
    \includegraphics[width=0.45\textwidth]{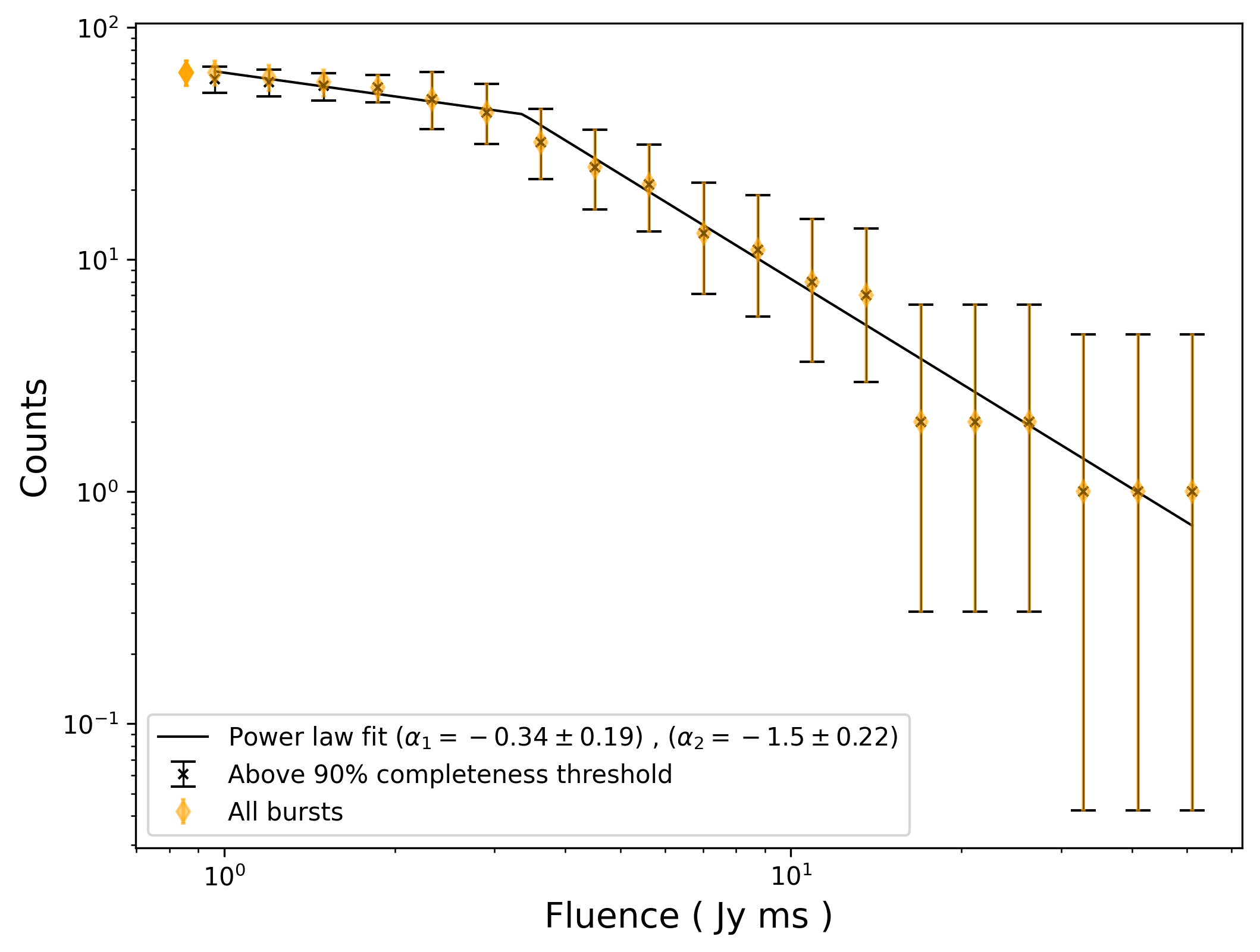}
    \caption{Same as Figure~\ref{fig:energydistribution_singleband} but for observations on 2023 August 08 (MJD 60164).} 
    \label{fig:energydistforindiepochs4}
\end{figure*}

\begin{figure*}
    \centering
    \includegraphics[width=0.45\textwidth]{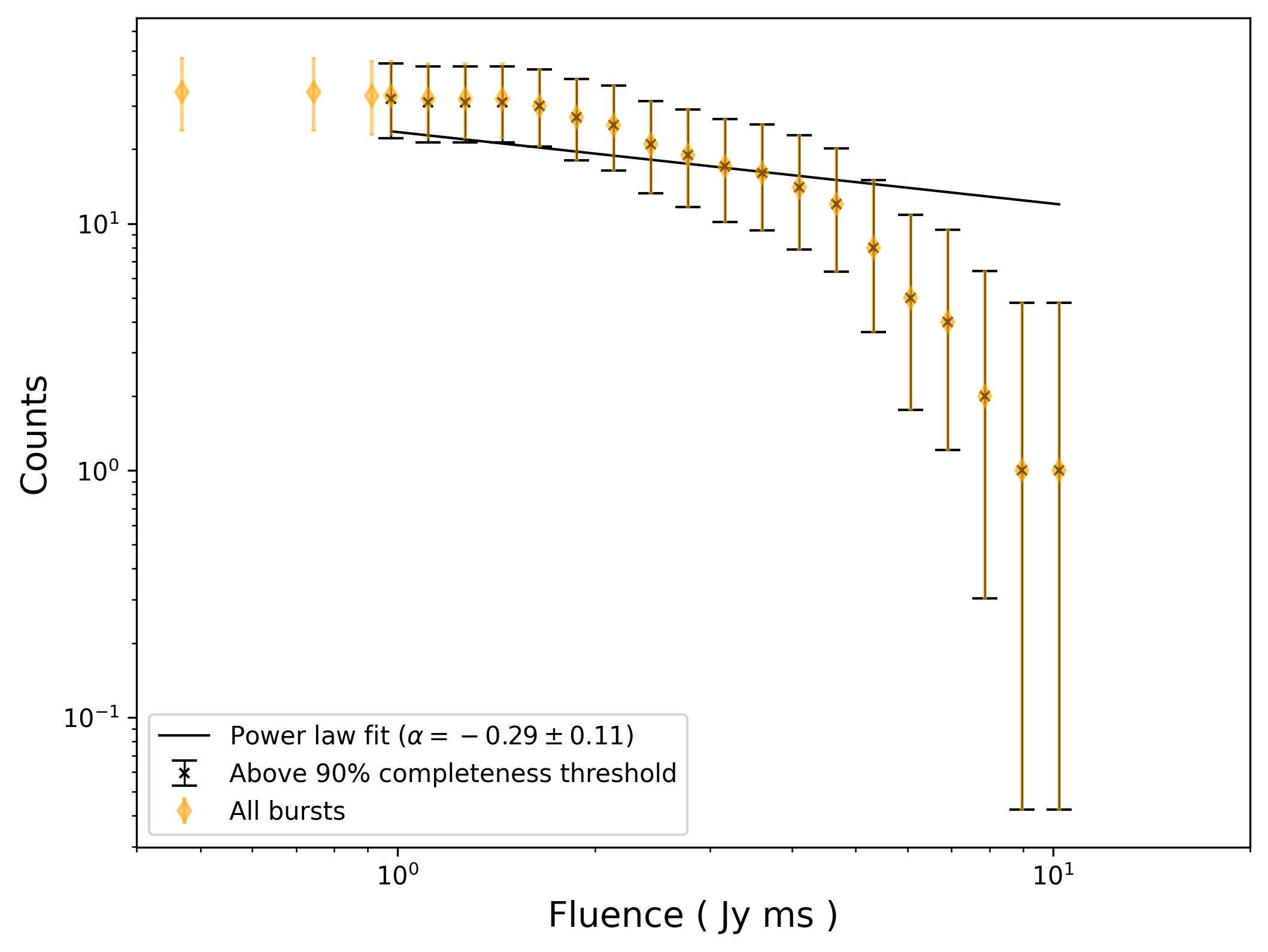}
    \hspace{0.2cm}
    \includegraphics[width=0.45\textwidth]{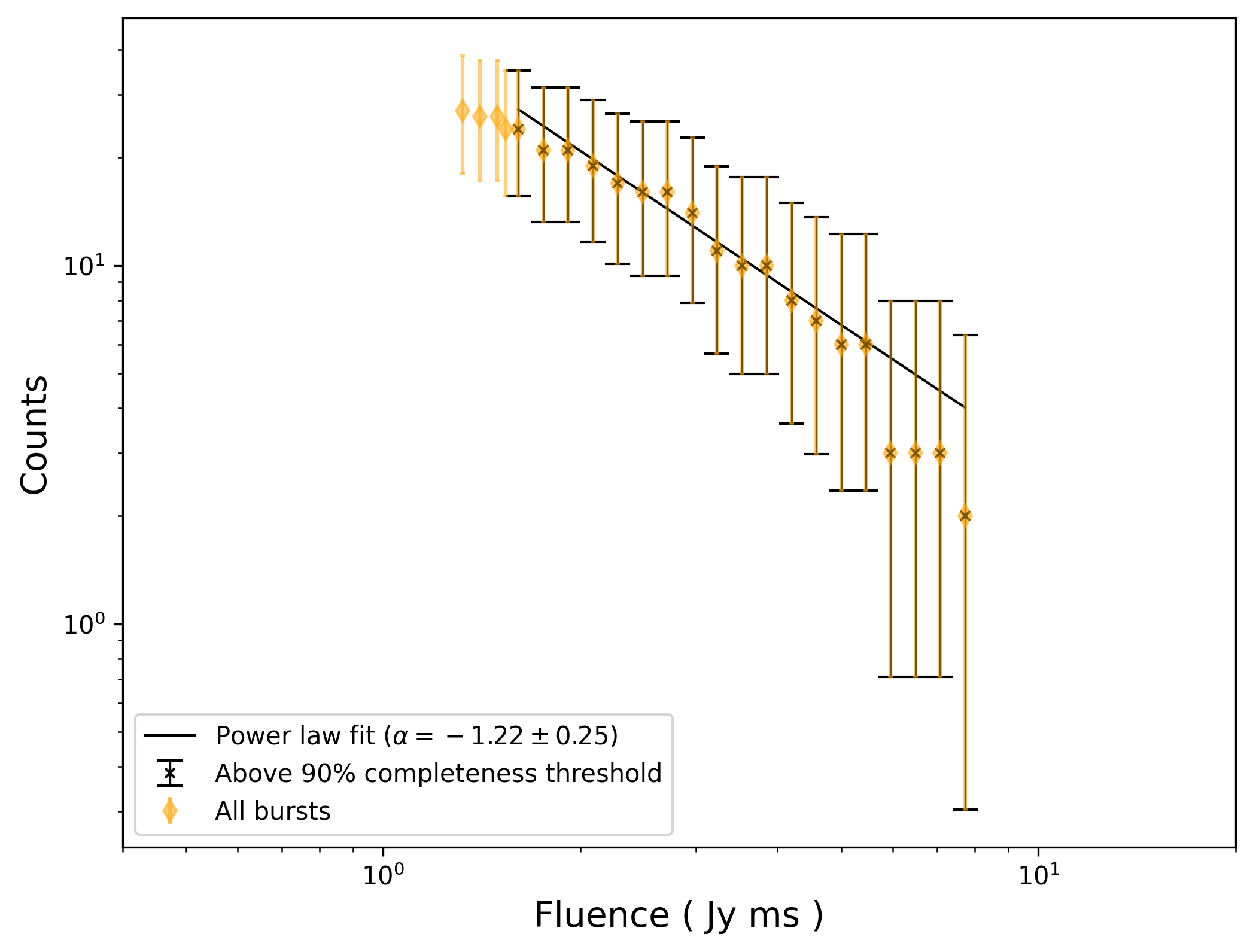}
    \caption{Same as Figure~\ref{fig:energydistribution_singleband} but for band-4 observations on \textit{left:} 2023 May 01 (MJD 60065)  and \textit{right:} 2023 August 10 (MJD 60166).} 
    \label{fig:energydistforindiepochs5}
\end{figure*}

\begin{figure*}
    \centering
    \includegraphics[width=0.7\textwidth, trim=0.5cm 4.5cm 0.5cm 4.5cm,clip]{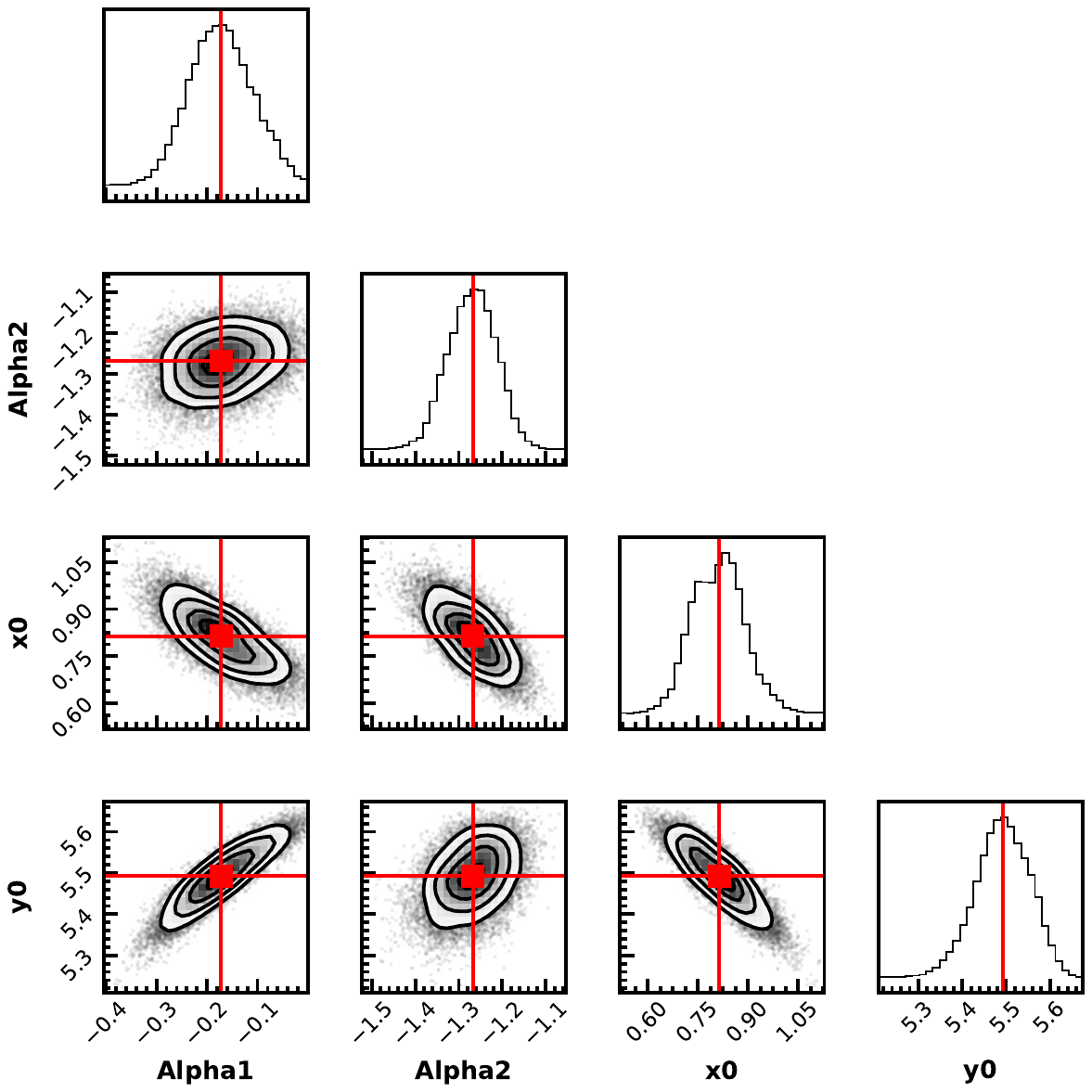}
    \caption{An example corner plot for fitting the cumulative distribution with MCMC. This formulation is shown for all the bursts detected above 90\% completeness threshold in band-4 over the whole observing campaign. Here x0 and y0 correspond to F$_{o}$ and N$_{o}$ described in the section~\ref{sec:plots_for_ind_epochs}.}  
    \label{fig:corner_plot_mcmc}
\end{figure*}

{
\noindent The cumulative fluence distribution of bursts is modelled using a broken power law of the form:

\begin{equation}
N(>F) =
\begin{cases}
N_0 \left(\dfrac{F}{F_0}\right)^{\alpha_1} & \text{if } F < F_0 \\[10pt]
N_0 \left(\dfrac{F}{F_0}\right)^{\alpha_2} & \text{if } F \geq F_0
\end{cases}
\label{eq:bpl}
\end{equation}

\noindent where $N(>F)$ is the cumulative number of bursts with fluence exceeding $F$, $\alpha_1$ and $\alpha_2$ are the power-law indices below and above the break, respectively, and $(F_0, N_0)$ defines the break point.

The fitting is performed in log-log space, where Equation~\ref{eq:bpl} reduces to two line segments. Specifically, for $\log F < \log F_0$:

\begin{equation}
\log N = \log N_0 + \alpha_1 \left(\log F - \log F_0\right),
\end{equation}

\noindent and for $\log F \geq \log F_0$:

\begin{equation}
\log N = \log N_0 + \alpha_2 \left(\log F - \log F_0\right).
\end{equation}

The model parameters $\boldsymbol{\theta} = \{\alpha_1,\, \alpha_2,\, 
\log F_0,\, \log N_0\}$ are estimated via Markov Chain Monte Carlo (MCMC) 
sampling using the \texttt{emcee} package \citep{Foreman-Mackey_2013}. 
The total variance on each data point is taken to be $\sigma_i^2$, 
where $\sigma_i$ is the quoted uncertainty. The log-likelihood function is:

\begin{equation}
\ln \mathcal{L}(\boldsymbol{\theta}) =
-\frac{1}{2} \sum_{i} \left[
\frac{\left(y_i^{\rm obs} - y_i^{\rm model}\right)^2}{\sigma_i^2}
+ 2\ln\left(\sigma_i\right)
\right],
\end{equation}

\noindent where $y_i^{\rm obs}$ and $y_i^{\rm model}$ are the observed and 
model predicted values of $\log N$ at the $i$-th data point, respectively. Asymmetric Poisson uncertainties are adopted: the square root of the counts for bins with $\geq 30$ bursts, and asymmetric uncertainties for bins with fewer than 30 bursts. The best-fit parameter estimates obtained from 
\texttt{scipy.optimize.minimize} are used as the initial guesses for the MCMC walkers.

The resulting fit for all the individual epochs and bands are shown in Figures~\ref{fig:energydistribution_singleband} to \ref{fig:energydistforindiepochs5}. As an example, the posterior distributions for all model parameters, along with their pairwise covariances, are displayed in Figure~\ref{fig:corner_plot_mcmc} for the band-4 bursts.}

\begin{figure*}
    \centering
    \includegraphics[width=0.9\textwidth]{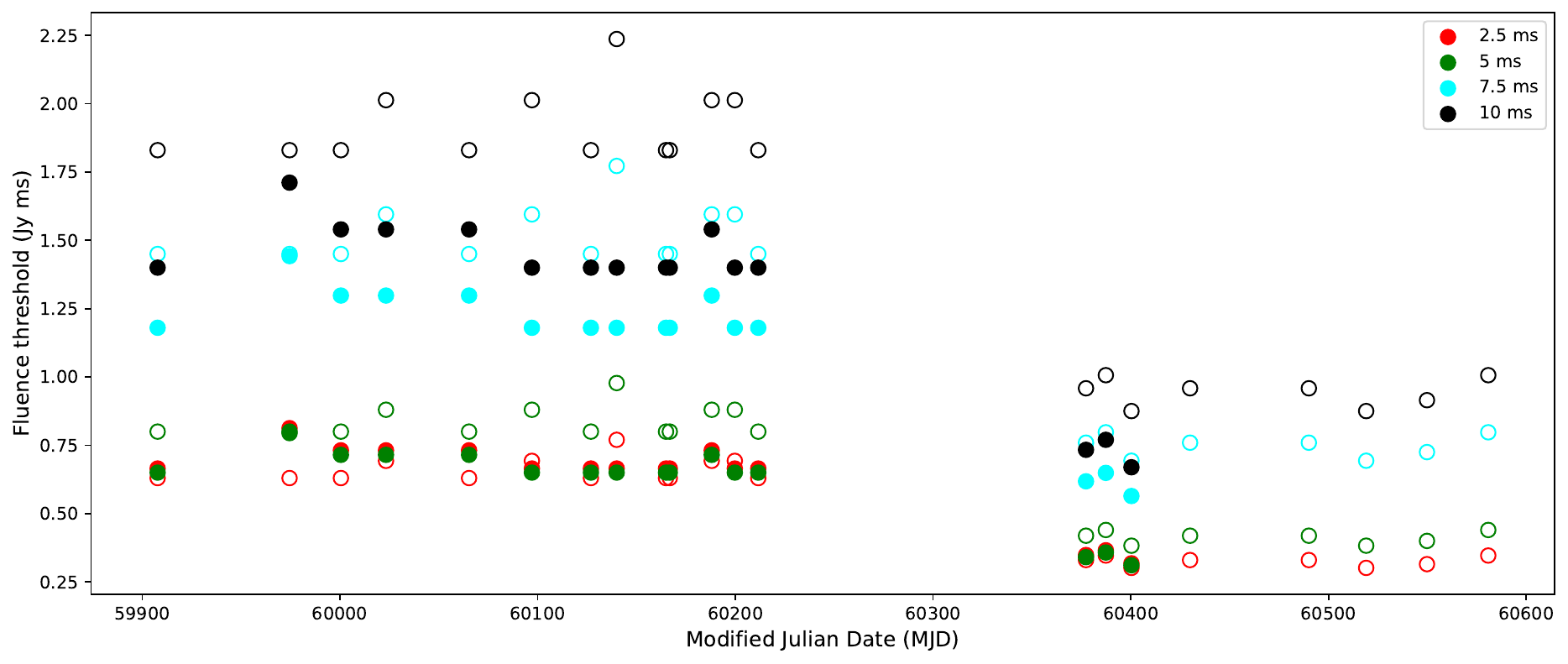}
    \caption{90\% completeness thresholds at different epochs of the observing campaign. Filled circles represent band-4 and hollow circles represent band-3. Different colors indicate the fluence threshold at different widths. The red and green points for 2.5 ms and 5 ms overlap for many epochs at band-4.}  
    \label{fig:compthresholds}
\end{figure*}

%

\end{document}